%% file: tc16_submit.tex
\documentclass[paper]{JHEP3} 
\usepackage{graphicx}
\usepackage{amssymb,amsfonts,amsmath}

\def\lsim{\mathrel{\rlap{\lower3pt\hbox{\hskip1pt$\sim$}}
     \raise1pt\hbox{$<$}}} 
\def\gsim{\mathrel{\rlap{\lower3pt\hbox{\hskip1pt$\sim$}}
     \raise1pt\hbox{$>$}}} 
\newcommand{\beq}{\begin{equation}}
\newcommand{\eeq}{\end{equation}}
\newcommand{\bea}{\begin{eqnarray}}
\newcommand{\eea}{\end{eqnarray}}

\title{Is there still any $T_c$ mystery in lattice QCD? \\
Results with physical masses in the continuum limit III}
\author{
Szabolcs~Bors\'{a}nyi$^a$\footnote{borsanyi@uni-wuppertal.de},
Zolt\'{a}n~Fodor$^{a,b,c}$, 
Christian Hoelbling$^a$,
S\'{a}ndor~D.~Katz$^c$,
Stefan~Krieg$^{a,d}$,
Claudia Ratti$^a$
and
K\'alm\'an~K.~Szab\'o$^a$\\
$^a$Department of Physics, University of Wuppertal, Gau\ss str. 20, D-42119, 
Germany\\
$^b$John von Neumann Institute for Computing, J\"ulich, D-52425, Germany\\
$^c$Institute for Theoretical Physics, E\"otv\"os University, P\'azm\'any
1, H-1117 Budapest, Hungary\\
$^d$Center for Theoretical Physics, MIT, Cambridge, MA 02139-4307, USA 
}

\abstract{
The present paper concludes our investigations on the QCD cross-over transition
temperatures with 2+1 staggered flavours and one-link stout improvement.  We
extend our previous two studies [Phys. Lett. B643 (2006) 46, JHEP 0906:088
(2009)] by choosing  even finer lattices ($N_t$=16) and we work again with
physical quark masses. The new results on this broad cross-over are in complete
agreement with our earlier ones. We compare our findings with the published
results of the hotQCD collaboration.  All these results are confronted with the
predictions of the Hadron Resonance Gas model and Chiral Perturbation Theory
for temperatures below the transition region. Our results can be reproduced by
using the physical spectrum in these analytic calculations. The findings of the
hotQCD collaboration can be recovered by using a distorted spectrum which takes
into account lattice discretization artifacts and heavier than physical quark
masses. This analysis provides a simple explanation for the observed
discrepancy in the transition temperatures between our and the hotQCD
collaborations. 
}
\preprint{WUB/10-11\\MIT-CTP-4152}

\begin{document}
\section{Introduction}
In recent years, increasing attention has been devoted to study the properties of the 
QCD phase diagram and thermodynamics. On the one hand, the heavy ion collision 
experiments at CERN SPS, RHIC at Brookhaven National Laboratory and ALICE 
at the Large Hadron Collider (LHC) provide the unique possibility of quantifying 
the properties of the deconfined phase of QCD. On the other hand, lattice calculations on QCD thermodynamics are reaching unprecedented levels of accuracy, with 
simulations at the physical quark masses and several values of the lattice cutoff: 
this allows to keep lattice artifacts under control. The 
information that can be obtained from these complementary approaches will shed 
light on the features of QCD matter under extreme conditions, one of the major 
challenges of the physics of strong interaction. 

One of the most interesting quantities that can be extracted from lattice
simulations is the transition temperature $T_c$ at which hadronic matter is
supposed to undergo a transition to a deconfined, quark-gluon phase. This
quantity has been vastly debated over the last few years, due to the
disagreement on its numerical value observed by different lattice
collaborations, which in some cases is as high as 20\% of the absolute value.
Indeed, the analysis of the hotQCD collaboration (performed with two different
improved staggered fermion actions, asqtad and p4, and with physical strange
quark mass and somewhat larger than physical $u$ and $d$ quark masses, $m_s
/m_{u,d} = 10$), indicates that the transition region lies in the range $T =
(185-195)$ MeV. Different observables lead to the same value of $T_c$ \cite{1,
2, Karsch:2007dt, Karsch:2008fe, Bazavov:2009zn}. 
Recent simulations using the p4 action with the quark mass ratio $m_s/m_u=20$
yielded about 5~MeV shift (towards the smaller values) in the temperature
dependence of the studied observables \cite{Cheng:2009zi}.
On the other hand, the results obtained
by our collaboration using the staggered stout action (with physical light and
strange quark masses, thus $m_s/m_{u,d}\simeq28$) are quite different: the
value of the transition temperature lies in the range 150-170 MeV, and it
changes with the observable used to define it \cite{6, 7}. This is not
surprising, since the transition is a cross-over \cite{8}: in this case it is
possible to speak about a transition region, in which different observables may
have their characteristic points at different temperature values, and the
temperature dependences of the various observables play a more important role
than any single $T_c$ value.  Unfortunately, the 25-30 MeV discrepancy was
observed between the two collaborations for the $T$ dependences of the various
observables, too.

A lot of effort has been invested, in order to find the origin of the
discrepancy between the results of the two collaborations.\footnote{ Note, that
quite recently preliminary results were presented \cite{20,Bazavov:2010sb} and
the results of the hotQCD collaboration moved closer to our results. (We
include some of these data in our comparisons.)} In Refs.
\cite{6,7}, we emphasized the role of the proper continuum limit
with physical quark masses, showing how the lack of them can distort the
result.  In \cite{Fodor:2007sy} we pointed out that the continuum limit can be
approached only if one reduces the unphysical pion splitting (the main
motivation of our choice of action).  An interesting application of these
observations was studied in \cite{Huovinen:2009yb}. These authors have
performed an analysis of trace anomaly, strangeness and baryon number
fluctuations within the Hadron Resonance Gas model (HRG).  They show that, to
reproduce the lattice results for the asqtad and p4 actions of the hotQCD
collaboration, it is necessary to distort the resonance spectrum away from the
physical one in order to take into account the larger quark masses used in
these lattice calculations, as well as finite lattice spacing effects.  As we
will see, no such distortion is needed to describe our data, and the
discrepancy between the two collaborations has its roots in the above mentioned
lattice artifacts.  In the present paper we perform a similar analysis for
those quantities that can be calculated in the HRG model and Chiral
Perturbation Theory ($\chi$PT), namely the chiral condensate, the strange quark
susceptibility and the equation of state. From the lattice point of view, we
present our most recent results for several physical quantities: our previous
calculations \cite{6, 7} have been extended to an even smaller lattice spacing
(down to $a \lsim0.075$ fm in the transition region), corresponding to $N_t =
16$. We use physical light and strange quark masses: we fix them by reproducing
$f_K/m_\pi$ and $f_K/m_K$ and by this procedure \cite{7} we get $m_s /m_{u,d} = 28.15$.
The HRG model results are obtained both for the physical resonance masses, as
listed in the Particle Data Book, and for the distorted spectrum which
corresponds to the quark masses and finite lattice spacings of
\cite{Bazavov:2009zn}. Our
analysis indicates that the discretization effects on hadron masses (and in
particular on the nondegenerate, taste-split light pseudoscalar meson masses
which emerge as a consequence of the staggered formalism) affect more severely
the asqtad and p4 actions than the stout one, in the temperature regime below and around
$T_c$. Indeed, the lattice results obtained with the stout action show a very
good agreement with the HRG model results with physical quark masses, while the
lattice results obtained with the asqtad and p4 actions can be reproduced
within the HRG model only with the distorted spectrum. The discrepancy in the transition
temperature values obtained by the two collaborations can be easily explained
by this result. 

The paper is organized as follows. In Section \ref{qcd} we give a brief review of the qualitative features of the QCD transition (those who are interested more in the qualitative features than in the technicalities might read this section and then jump directly to subsection \ref{results}). In Section \ref{lattice} we give the details of our numerical simulations. In Section \ref{latresults} we present the results of our simulations for different observables. In Section \ref{hrg} we present some aspects of the Hadron Resonance Gas 
model and the comparison 
between lattice and HRG model results. We write our Conclusions in Section \ref{conclusions}. 
In Appendix A we provide some details of the chiral condensate calculation 
in the HRG model + $\chi$PT. Appendix B presents the temperature dependence of
our continuum extrapolated lattice results.

\section{The QCD transition \label{qcd}}

In this section we summarize the qualitative features of the $T>0$ QCD 
transition. One of the most important pieces of information we have is 
our knowledge about the nature of the transition. Though many take it 
for granted, it is a higly non-trivial result, that the transition is an 
analytic one and usually called a cross-over \cite{8}. In 
order to show this by means of lattice QCD, physical quark masses were 
taken, and a finite size scaling analysis was carried out for the 
continuum extrapolated chiral susceptibilities. This analytic behaviour 
has important consequences for any $T_c$ determination in QCD.

In order to illustrate the most important differences between a real 
phase transition and an analytic cross-over, we recall the water-vapor 
phase diagram on the temperature versus pressure plane (c.f. 
\cite{6} and Figure \ref{fig:steam} of the present paper). We study the transition by fixing the pressure to a 
given value and then varying the temperature. For smaller pressures 
(p$\lsim$22~MPa) there is a first order phase transition. The density 
jumps, the heat capacity is infinite, and these singular features appear 
simultenously, thus exactly at the same critical temperature. At 
pressure $p\approx$22.064~MPa and temperature $T\approx$647.096~K, there is 
a critical point with a second order phase transition. This phase 
transition is also characterized by a singular behaviour. \footnote{Note, 
that a 
real singularity, a phase transition, takes place only in infinite size 
systems. In our example we have a macroscopic amount of water with {\cal 
O}(10$^{23}$) molecules.  From the practical point of view, this is an 
infinitely large system.}

%\begin{figure}[h!]
\FIGURE{
\centerline{\includegraphics*[width=9cm,viewport=20 170 590 630,clip]{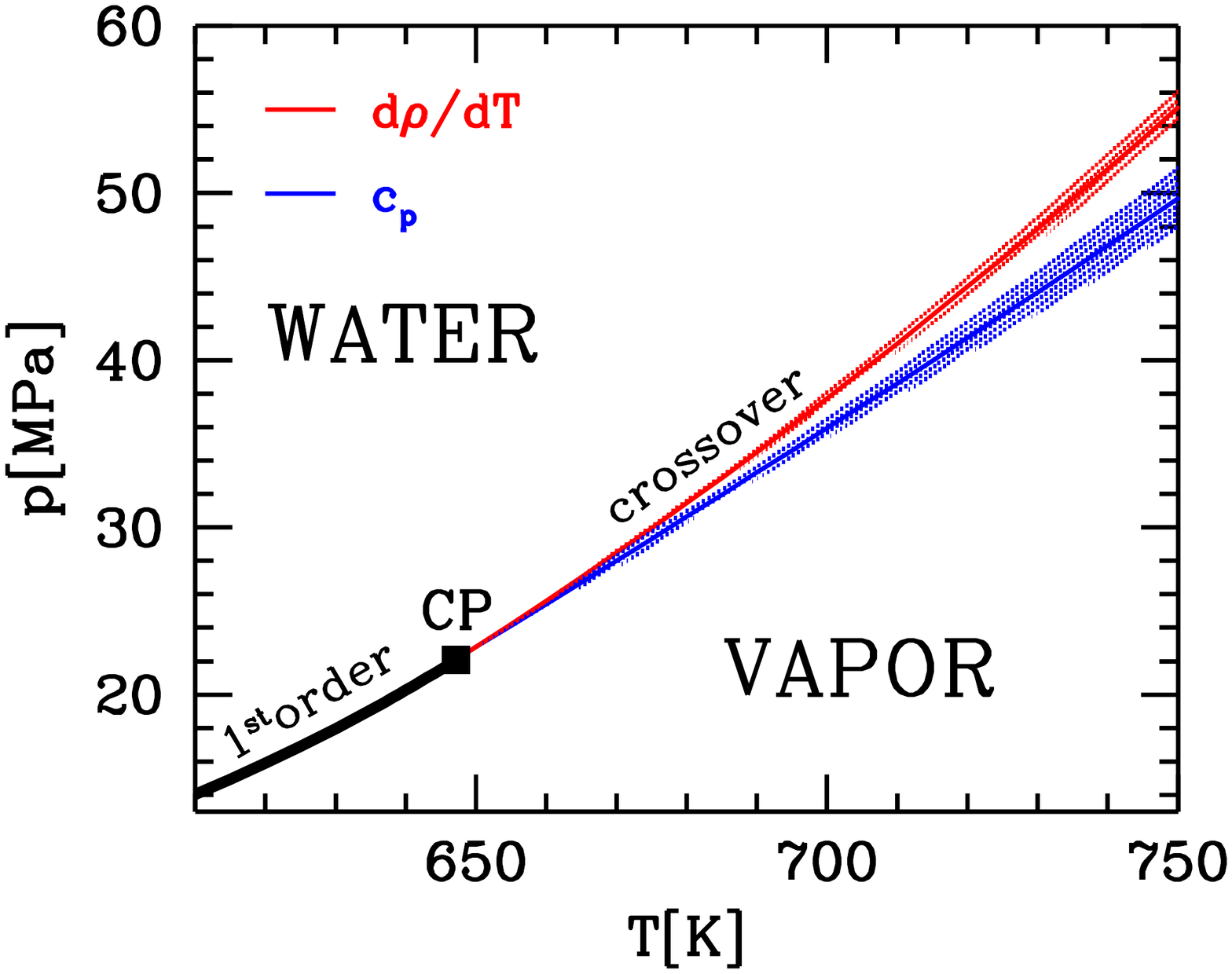}}
\caption{\label{fig:steam}
The phase diagram of water around its critical point (CP). For pressures below the critical
value ($p_c$) the transition is first order, for $p>p_c$ values there is a rapid cross-over.
In the cross-over region the critical temperatures defined from different quantities are not
necessarily equal. This can be seen for the temperature derivative of the density ($d\rho/dT$) and
the specific heat ($c_p$). The bands show the non-negligible 
experimental uncertainties (see \cite{Spang}).
}
}
%\end{figure}

At even larger pressures (p$\gsim$22.064 MPa) the water-vapor transition 
is an analytic one (the behaviours of various observables are analytic, 
even in the infinite volume limit). As a consequence, in this pressure 
region there is no jump in the density when we change the temperature, only 
a rapid but continuous change. The inflection point of this 
density-temperature function (the point with the largest, though finite, 
derivative) can be used to define the pseudocritical temperature 
(another usual name for it is ``transition temperature") related to the 
density. Similarly, the heat capacity is always finite, but it has a 
pronounced peak as we increase the temperature. The position of this 
peak can be used to define the pseudocritical temperature related to the 
heat capacity. Despite the fact that there is no singularity, the 
inflection point and peak position are well defined. The corresponding
pseudocritical or transition temperature is usually denoted as $T_c$.

The most important message here is that the various transition 
temperatures (e.g. those related to the density or heat capacity) behave 
differently depending on whether we are in the singular (real phase 
transition) or non-singular (analytic cross-over) region. As it is 
indicated on the figure, for a real phase transition these critical 
temperatures coincide, whereas in the non-singular region (for pressures 
above 22.064 MPa) the pseudocritical temperatures can differ 
considerably. The fast change (though no jump) in the density is at a 
lower temperature than the peak in the heat capacity. The transition is 
a broad cross-over. The pseudocritical temperatures, related to various 
observables, are separated, but both of them are in the broad transition 
temperature region. This separation does not mean that we have two 
transitions (one for the density and one for the heat capacity), it 
merely reflects the broadness of the transition.

It is easy to see that different observables can give different 
pseudocritical temperatures. Let us study an observable $X$, which 
characterizes the transition as a function of the temperature $X(T)$. 
For a real phase transition its singular behaviour appears at the same 
temperature even if we multiplied it by $T$ (an infinitely high peak  
keeps its position). For an analytic cross-over, we have a peak with a finite 
height and a finite width. Multiplying it by $T$ shifts the peak 
position to larger temperature values. The value of $T_c$ is shifted. 
The pseudocritical temperature is well defined for any definition, but 
it is not unique. Furthermore, for a broad transition the whole
neighbourhood of the peak behaves similarly as the peak, the determination
of the peak's or inflexion point's position is difficult 
(this is the experimental reason for the uncertainties on
Figure \ref{fig:steam} and this technical difficulty is present for the even broader QCD transition). Though a $T_c$ related to some observable is 
informative, a more complete description is given by the whole 
temperature dependence of $X(T)$.

The determination of such curves is the main goal of any study on the 
QCD transition (c.f. our earlier studies 
\cite{6,7}). Since the QCD transition (at vanishing 
chemical potential) is an analytic cross-over, one wants to obtain these 
smooth curves for several observables. Though the characteristic points 
of such curves contain obviously less information than the curves 
themselves, we give them, too.

Before we list the observables we study in detail, it is worth 
mentioning that the cross-over nature of the QCD transition is related to 
the specific values of the quark masses we have in nature. For two- or 
three-flavour QCD with vanishing quark masses or with infinitely massive 
quarks, one is supposed to have real phase transitions. There are order parameters (in 
the former case the chiral susceptibility/condensate signaling the 
chiral phase transition; in the latter case the Polyakov line signaling 
the deconfinement phase transition) which show a non-analytic behaviour 
as we change the temperature. As we pointed out earlier, the highly 
non-trivial result about the analytic nature of the QCD transition with 
physical quark masses implies, that no observable can be treated as an 
order parameter. All of the observables show analytic temperature 
dependences. There is neither a chiral nor a deconfinement phase 
transition. Note however, that similarly to the density or to the heat 
capacity in the water-vapor cross-over transition, the observables 
chiral susceptibility/condensate and the Polyakov line can develop a 
pronounced peak or show a rapid change. The peak positions or the 
inflection points for such a cross-over are usually expected to be at 
different temperatures. Again, we do not say 
\cite{6,7} that there are two phase transitions and 
one of them is at a lower temperature than the other. The separation of 
the pseudocritical temperatures is merely a sign of the broad analytic 
transition \cite{8}.

Since the chiral susceptibility/condensate and the Polyakov loop are not 
order parameters, they are just used to signal the cross-over. In 
principle any other quantity showing rapid changes or developing a peak 
in the transition region can be studied. The temperature dependences of 
these observables can be compared with the predictions of other lattice 
results or model calculations. In this paper we extend our analysis to 
new observables and to even finer lattices. We study the above 
chiral/deconfinement observables and in addition we look at the strange quark number 
susceptibility and at the energy density or trace anomaly \cite{EOS}.

The reason for calculating the temperature dependence of these many 
observables is obvious. The more observables we study, the broader 
picture we have on the QCD transition. To be more specific, the chiral 
susceptibility/condensate and the Polyakov loop are remnants of the real 
phase transition order parameters (for other mass regions of the phase 
plane). In addition to our old observables we use a new definition for 
the chiral condensate, which has adventageous renormalization features 
and gives a result with little noise (due to construction, the chiral 
susceptibility is somewhat noisy). The strange quark number 
susceptibility is a particularly attractive quantity from the 
theoretical point of view. It is related to a conserved current, thus no 
renormalization ambiguities appear, which makes direct comparisons 
particularly easy. For a first order phase transition, the energy density 
has a jump. In the cross-over region the remnant of this jump is an 
inflection point. Furthermore, the transition temperature related to the 
equation of state has a direct link to experiments, 
its importance is obvious.

The various obsevables (listed in the previous paragraph) lead to 
different transition temperatures, they are typically between 150 and 
170 MeV, thus well within the broadness of the transition. Let us 
emphasize again, the difference between the pseudocritical
$T_c$ values does not mean 
that one of the phase transitions happens at a lower temperature than the 
other, quite the contrary: no phase transition happens at all. Our new 
results confirm our earlier findings and their interpretation by all 
means: the transition temperatures scatter within the broad temperature 
interval, characteristic of the cross-over.

\section{Details of the lattice simulations \label{lattice}}
\subsection{Action, algorithm and scale setting}
The lattice action is the same as we used in \cite{6, 7}, namely a tree-level Symanzik improved gauge, and a stout-improved staggered fermionic action (see Ref. \cite{Aoki:2005vt} for details).  The stout-smearing 
\cite{Morningstar:2003gk} reduces the taste violation (see Section \ref{taste}): this kind of smearing has the 
smallest taste violation among the ones used so far in the literature for large
scale thermodynamical simulations\footnote{Only recently, first exploratory
studies of the hotQCD Collaboration with the HISQ action
\cite{20,Bazavov:2010sb} start to
appear: in this case, the projected smeared links improve the taste symmetry in
a similar way as in our stout action.}. The supression of this
artefact is important in the transition region (see the important consequences
within the hadron resonance model framework) and that was the main motivation
for this choice. For details about the algorithm we refer the reader to 
\cite{7}. 

In analogy with what we did in \cite{6,7}, we set the scale at the physical 
point by simulating at $T=0$ with physical quark masses \cite{7} and reproducing the kaon and pion masses and the kaon decay constant. This gives 
an  uncertainty of about 2\% in the scale setting, which propagates in the uncertainty in the 
determination of the temperature values listed.
\subsection{Taste violation \label{taste}}
Most of the large scale QCD thermodynamics studies apply the staggered 
formalism for the quark fields. Working at non-vanishing lattice spacing 
within this framework, there is only one single pseudo-Goldstone boson 
(instead of the experimentally observed three pions). By 
pseudo-Goldstone we mean a particle whose mass approaches zero if we tune 
the mass of the quark to zero (note, that in lattice studies usually we 
do not approach this zero mass --chiral-- limit, but we tune the quark 
masses to their physical values). In addition to this single 
pseudo-Goldstone boson, there is a whole tower of non-Goldstones. They are 
usually much heavier, which is a non-physical lattice artifact. The typical 
mass gap can be as large as several hundred MeV, which vanishes as the 
lattice spacing tends to zero and one recovers the experimentally 
observed spectrum (since the original staggered formulation provides four 
flavours --or how they are called in lattice QCD: tastes-- the proper number 
of degrees of freedom is reached by taking the root of the fermion 
determinant). On the more formal level, this implies that every pseudoscalar meson (for example pions and 
kaons) is split into 16 non-degenerate mesons, which can be grouped into the eight 
multiplets \cite{12} listed in Table \ref{table1}. Their masses can be written as:
\beq
m^{2}_{i} = m^{2}_{0} + (\delta m_i )^2 .
\label{delta}
\eeq 

\TABLE{
\begin{tabular}{c|c|c}
index&$\Gamma^F$&multiplicity $n_i$\\
\hline
0&$\gamma_5$&1\\
1&$\gamma_0\gamma_5$&1\\
2&$\gamma_i\gamma_5$&3\\
3&$\gamma_i\gamma_j$&3\\
4&$\gamma_i\gamma_0$&3\\
5&$\gamma_i$&3\\
6&$\gamma_0$&1\\
7&1&1\\
\hline
\end{tabular}
\caption{Left column: index shown in Fig. \ref{fig1}. Central column: taste matrices $\Gamma^F$ . 
Right column: multiplicity of the different pseudoscalar mesons. 
}
\label{table1}
}

The splittings $ (\delta m_i )^2$ are proportional to $(\alpha_s a^2 )$ for small lattice spacings. Only one out of the 16 pseudoscalar mesons is a true Goldstone boson in the chiral limit. The splitting  
(the taste symmetry violation) has to vanish in the continuum limit. Once it 
shows an $\alpha_s a^2$ dependence (in practice a quadratic dependence
with a subdominant logarithmic correction) we are in the scaling region.  
This is an important 
check for the validity of the staggered framework at a given
lattice spacing (for large lattice spacings its behaviour can mimic an
incorrect continuum limit). In Ref. \cite{7} we showed a 
continuum extrapolation of the quadratic mass difference $(\delta m_i )^2$, concluding that 
the splitting obtained with the stout action is consistent with zero in the continuum 
limit. We also showed that lattice spacings which are larger than $a \sim 0.15$ fm are 
not in the expected $a^2$-scaling regime. In Fig. \ref{fig1} we show the leading order $a^2$-behavior of the 
masses of the pion multiplets calculated with the asqtad (left panel) and stout (right 
panel) actions. It is evident that the 
continuum expectation is reached faster in the stout action than in the asqtad one. In addition, 
in the present 
paper we push our results to $N_t = 16$, which corresponds to even smaller lattice 
spacings and mass splittings than those used in \cite{7}. From Fig. \ref{fig1}, we can obtain the lattice spacing-dependent 
spectrum that we will include in the HRG model, in order to take into account 
lattice discretization effects.

\FIGURE{
%\begin{figure}
\hspace{-.8cm}
\begin{minipage}{.48\textwidth}
\parbox{6cm}{
\scalebox{.71}{
\includegraphics{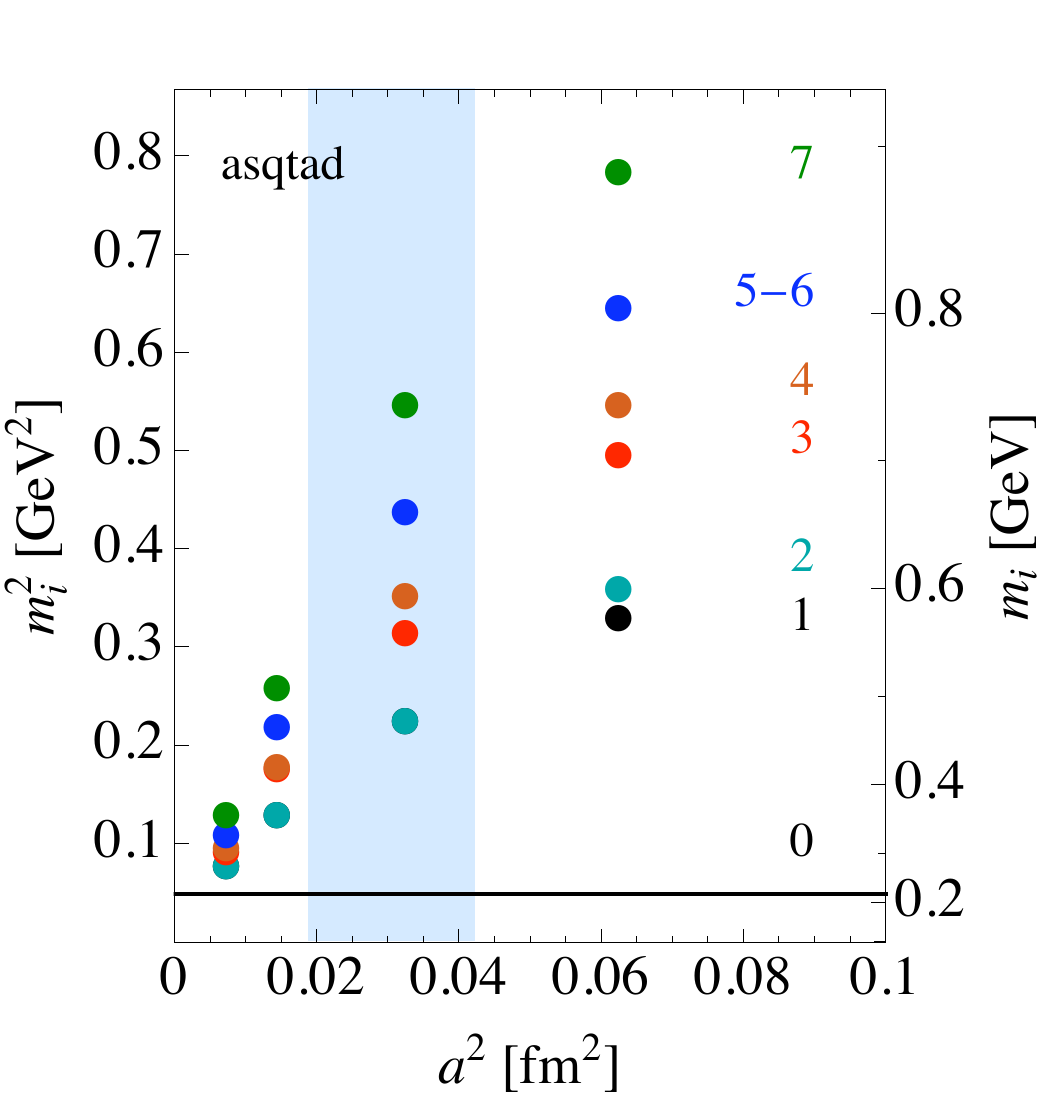}\\}}
%\centerline{(a)}
\end{minipage}
\hspace{.4cm}
\begin{minipage}{.48\textwidth}
\hspace{.4cm}
\parbox{6cm}{
\scalebox{.71}{
\includegraphics{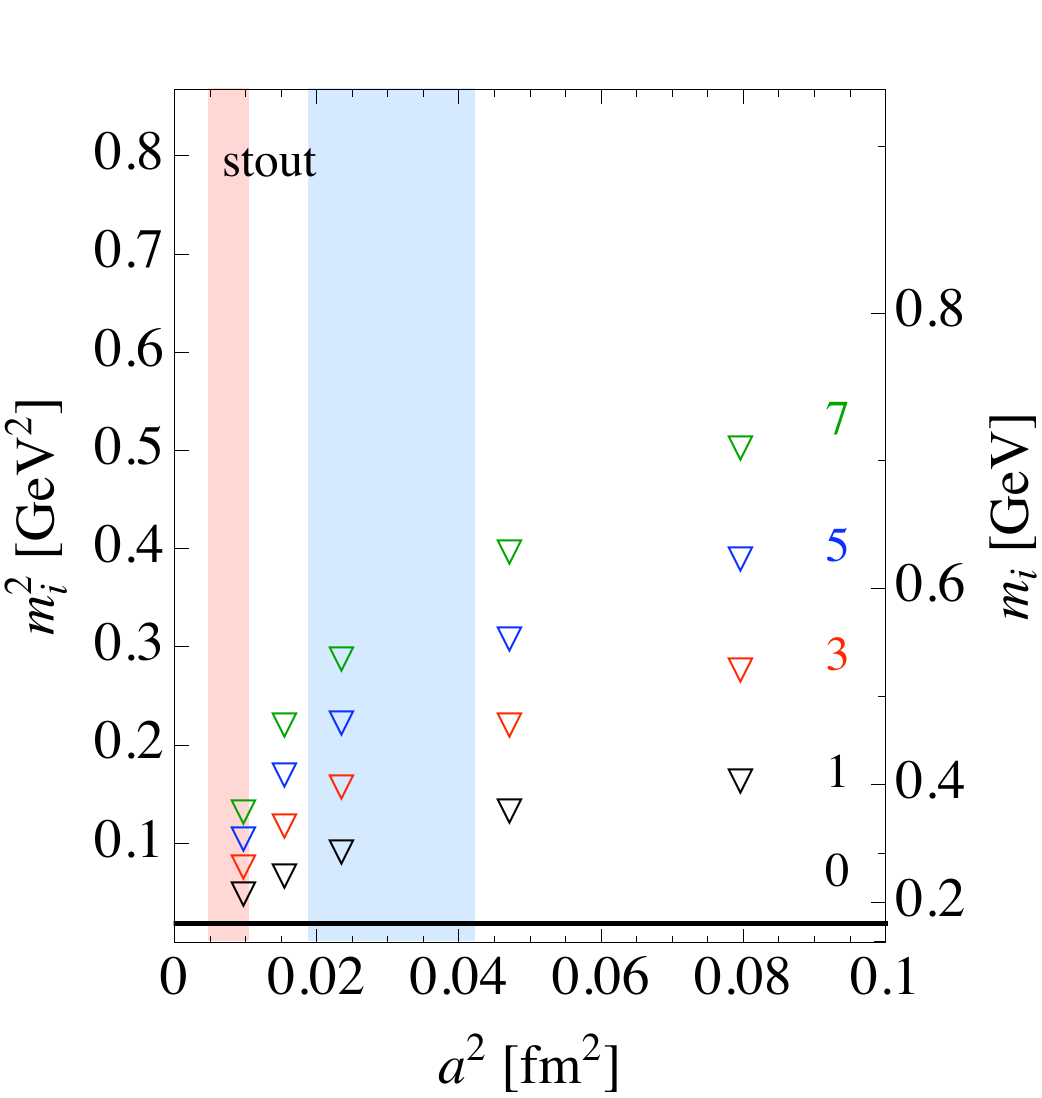}\\}}
%\centerline{(b)}
\end{minipage}
\caption{Masses of the pion multiplet squared, as functions of the lattice spacing squared. 
Left panel: asqtad action \cite{Bazavov:2009bb}. Right panel: stout action. The numbers
next to the data correspond to the taste matrices, as listed in Table
\ref{table1}. In both panels, the blue band indicates the relevant range of
lattice spacings for a thermodynamics study at $N_t=8$ between $T=$ 120 and 180
MeV. The red band in the right panel corresponds to the same temperature range
and $N_t=16$. In both figures, the horizontal line labelled as ``0" is the
pseudo-Goldstone boson, which has a mass of 220 MeV for the asqtad results, and
135 MeV for the stout ones (As we mentioned the splitting is formally
proportional to $\alpha_s a^2$.  At present accuracies and for illustrative
purposes the subdominant logarithmic dependence can be omitted).
}
\label{fig1}
%\end{figure}
}

\section{Lattice results\label{latresults}}
In this Section we present our lattice results for the strange quark number
susceptibility, Polyakov loop and two different definitions of the chiral
condensate. After performing a continuum extrapolation, we extract the values
of the transition temperature associated to these observables. As we
alreay emphasized the temperature dependence of an observable
contains much more information than the location of a peak or
inflection point (which are usually hard to determine precisely for such a broad 
transition). 
We perform a HRG analysis and compare our results with those of the 
hotQCD Collaboration in the next Section. 

Quark number susceptibilities are defined in the following way:
\beq
\chi_{2}^{q}=\left.\frac{T}{V}\frac{\partial^2\ln Z}{\partial(\mu_q)^2}\right|_{\mu_i=0},
~~~~~~q=u,d,s.
\label{qns}
\eeq
These quantities rapidly increase during the transition, therefore
they can be used to identify this region. However, while light quark susceptibilities are
dominated by pions at small temperatures, kaons are the lightest degrees of
freedom for strange quark susceptibilities in the hadronic phase. Therefore,
these two quantities are known to behave very differently as functions of the
temperature, with the strange quark number susceptibility rising more slowly in
the transition region. Due to the presence of disconnected diagrams, the light
quark number susceptibility is known to be very noisy and was not calculated.
Nevertheless, we will discuss its temperature dependence within the hadron
resonance gas model in the next Section.
 
In the left panel of Fig. \ref{deco} we show our results for the strange quark
number susceptibility for $N_t=$ 10, 12, 16. The gray band is the continuum
extrapolation that we have performed using our data: the numerical values are
listed in the Table of Appendix \ref{B} (the width of this band and those for other observables indicate the statistical and systematic uncertainties of the continuum extrapolation). 

The Polyakov loop is the order parameter related to the deconfinement phase
transition of QCD in the pure gauge sector. In this case, the $Z_3$ symmetry is
exact at small temperatures, where the Polyakov loop expectation value is zero.
In the deconfined phase, this symmetry is spontaneously broken by the
expectation value of the Polyakov loop, which jumps to a finite value.  When
quarks are included in the system, the $Z_3$ symmetry is explicitly broken by
their presence. In this case, the Polyakov loop is no longer a real order
parameter. Nevertheless, it is still considered as an indicator for the
transition, since it exhibits a rise in the transition region.  This is evident
from the right panel of Fig. \ref{deco}, where we plot the renormalized
Polyakov loop as a function of the temperature. The need to renormalize it
comes from the fact that there are self-energy contributions to the static
quark free energy that need to be eliminated. To that end, we use our
renormalization procedure of \cite{6}. In order to compare our results with
those obtained by the hotQCD collaboration \cite{Bazavov:2009zn} (which will be done in the
next Section), the renormalization constant is obtained slightly differently
from the condition
$V(1.5 r_0)=V_{\rm string}(1.5r_0)$ where V is the zero temperature quark-antiquark
potential and $V_{\rm string}(r)=-\pi/12r+\sigma r$.
In addition, we included the factor $\frac 13$ in the trace
definition. 
%As an intermediate step towards the renormalized Polyakov loop $L(T)$
%we removed the divergences by using the  $L(T)/L^2(2T)$ combination, 
%where the enumerator and
%the denominator are measured at the same lattice spacing. We then multiplied 
%this by $L^2(2T)$ obtained from simulations with fixed lattice spacings 
%by varying $N_t=6\dots 12$.

The right panel of Figure \ref{deco} shows the different $N_t$ data sets together with the
continuum extrapolated result, for which we give numerical data in the Table
of Appendix \ref{B}.
As it is expected from a broad cross-over the rise of the Polyakov
loop is pretty slow as we increase the temperature (c.f.
\cite{Bazavov:2009zn,6,7}).

\FIGURE{
%\begin{figure}
\hspace{-.8cm}
\begin{minipage}{.48\textwidth}
\parbox{6cm}{
\scalebox{.68}{
\includegraphics{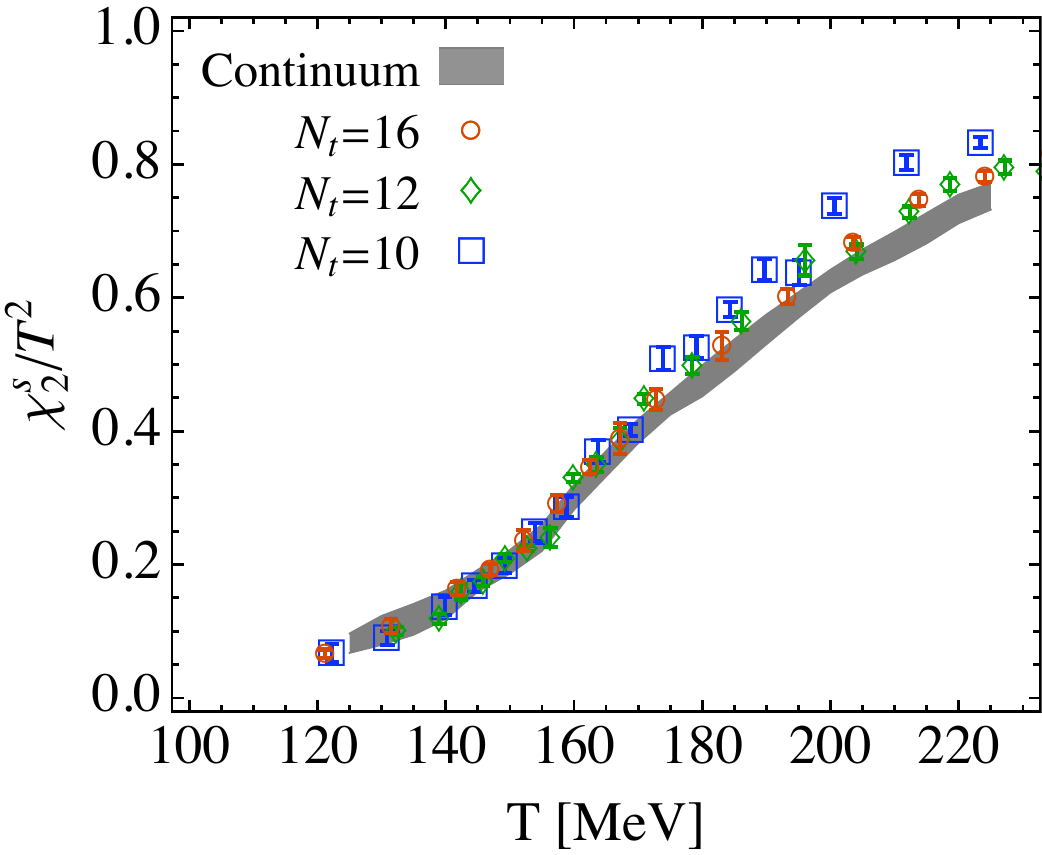}\\}}
%\centerline{(a)}
\end{minipage}
\hspace{.28cm}
\begin{minipage}{.48\textwidth}
\parbox{6cm}{
\scalebox{.68}{
\includegraphics{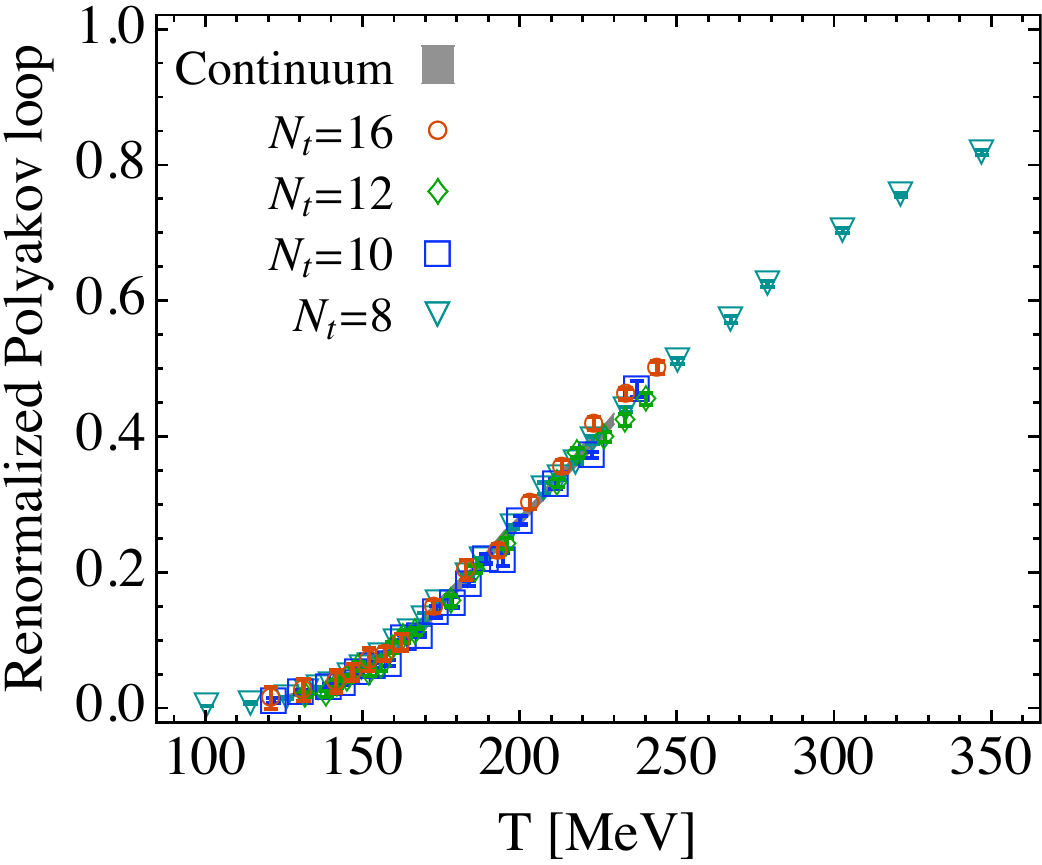}\\}}
%\centerline{(b)}
\end{minipage}
\caption{
Left: strange quark number susceptibility as a function of the temperature. Right: renormalized Polyakov loop as a function of the temperature. In both figures, the different symbols correspond to different $N_t$. The gray band is the continuum extrapolated result.}
\label{deco}
%\end{figure}
}

\FIGURE{
\hspace{-.8cm}
\begin{minipage}{.48\textwidth}
\parbox{6cm}{
\scalebox{.68}{
\includegraphics{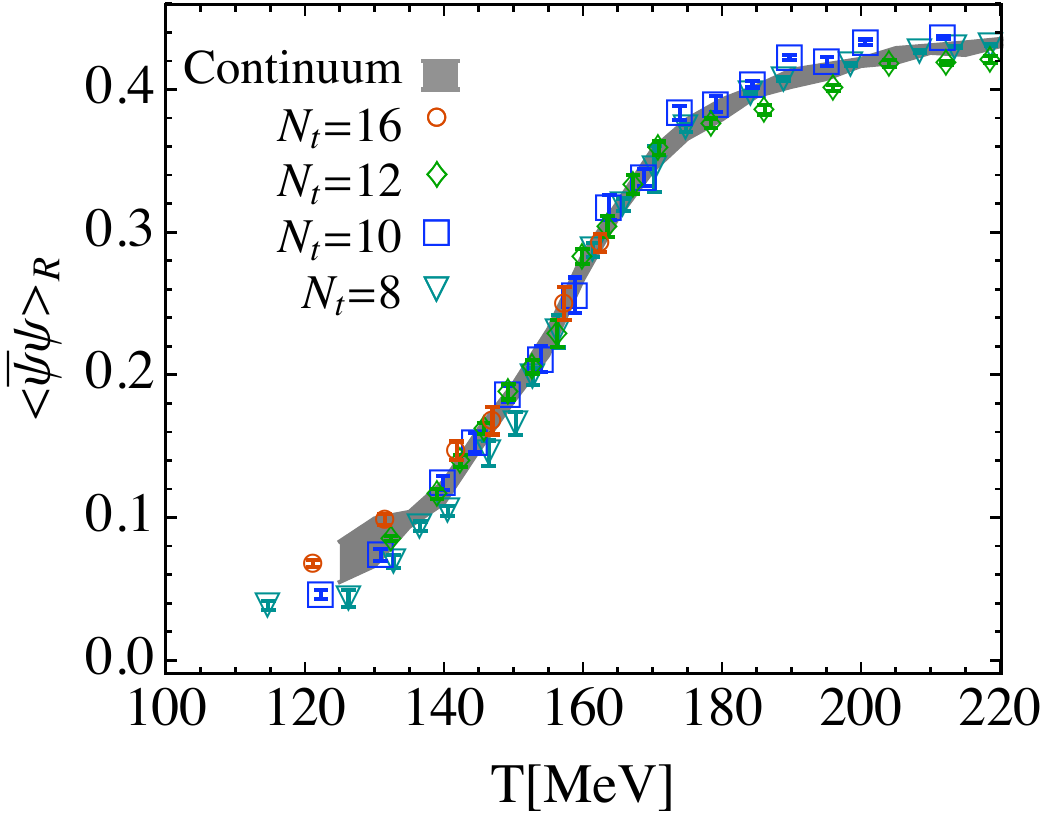}\\}}
%\centerline{(a)}
\end{minipage}
\hspace{.28cm}
\begin{minipage}{.48\textwidth}
\parbox{6cm}{
\scalebox{.68}{
\includegraphics{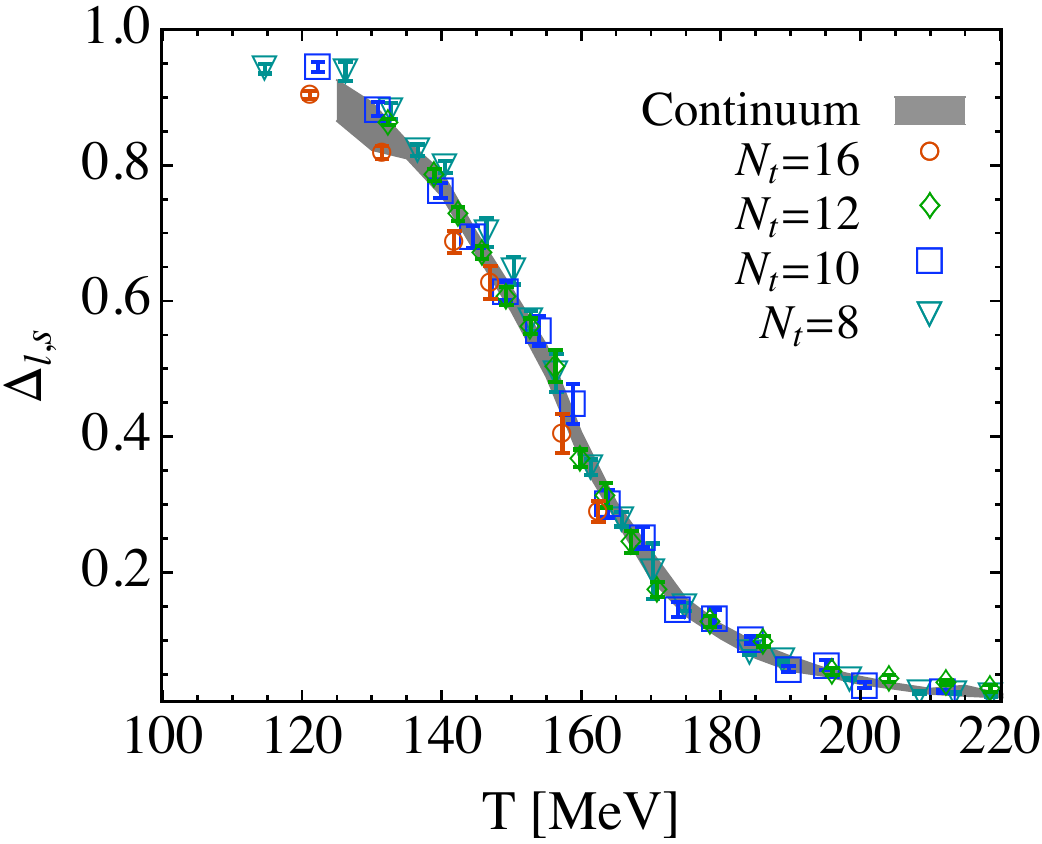}\\}}
%\centerline{(b)}
\end{minipage}
\caption{
Left: renormalized chiral condensate $\langle\bar{\psi}\psi\rangle_R$ defined
in Eq. (\ref{pbp}). Right: subtracted chiral condensate $\Delta_{l,s}$ defined
in Eq. (\ref{deltals}).  In both figures, the different symbols correspond to
different $N_t$. The gray band is our continuum estimate.}
\label{chir}
}
%\end{figure}

The chiral condensate is defined in the following way:
\beq
\langle\bar{\psi}\psi\rangle_q=\frac{T}{V}\frac{\partial\ln Z}{\partial m_q},~~~~~~q=u,d,s.
\eeq
In the case of a real chiral phase transition, the chiral condensate is the
corresponding order parameter.  However, with physical quark masses there is no
real phase transition, just a cross-over. The chiral condensate can still be
taken as an indicator for the remnant of the chiral transition, since it
rapidly changes in the transition region.

In the present paper, the following definition of the renormalized chiral condensate is used:
\beq
\langle\bar{\psi}\psi\rangle_R=-\left[\langle\bar{\psi}\psi\rangle_{l,T}-\langle\bar{\psi}\psi\rangle_{l,0}\right]\frac{m_l}{X^{4}}~~~~~~l=u,d.
\label{pbp}
\eeq
In the above equation, $X$ can be any quantity which has a dimension
of mass. Since we are working with non-vanishing quark masses, $m_\pi$ is a reasonable choice.
This quantity is properly renormalized and the continuum limit can be safely taken \cite{8}.
The individual results and the continuum extrapolation are shown 
in Figure~\ref{chir}.

In order to compare our results to those of the hotQCD collaboration, we also calculate the quantity
$\Delta_{l,s}$, which is defined as
\beq
\Delta_{l,s}=\frac{\langle\bar{\psi}\psi\rangle_{l,T}-\frac{m_l}{m_s}\langle\bar{\psi}\psi\rangle_{s,T}}
{\langle\bar{\psi}\psi\rangle_{l,0}-\frac{m_l}{m_s}\langle\bar{\psi}\psi\rangle_{s,0}}~~~~~~l=u,d.
\label{deltals}
\eeq
Since the results at different lattice spacings are essentially 
on top of each other, we connect
them to lead the eye and use this band in later comparisons (c.f. 
Fig. \ref{chir}). 

\TABLE{
%\begin{table}
\begin{tabular}{|c||c|c|c|c|c|c|}
\hline
&$\chi_{\bar\psi\psi}/T^4$&$\Delta_{l,s}$&$\langle\bar{\psi}\psi\rangle_R$&$\chi_{2}^{s}/T^2$&${\epsilon/T^4}$&$(\epsilon-3p)/T^4$\\
\hline
this work&147(2)(3)&157(3)(3)&155(3)(3)&165(5)(3)&157(4)(3)&154(4)(3)\\
our work '09&146(2)(3)&155(2)(3)&-&169(3)(3)&-&-\\
our work '06&151(3)(3)&-&-&175(2)(4)&-&-\\
\hline
\end{tabular}
\caption{The pseudocritical temperatures in MeV (defined as the inflection
point or peak position of the $T$ dependent observables listed in Section
\ref{qcd}) for physical quark masses in the continuum limit.  The $T_c$ values
from the equation of state (energy density and trace anomaly) are not continuum
extrapolated, they are obtained on $N_t=8$ lattices. We expect a shift within
the error bars in the continuum limit.  A comparison between our present and
earlier results \cite{6} and \cite{7} is given. A change in the experimental
$f_K$ value in 2008 resulted in a $\approx$6~MeV reduction of our $T_c$
predictions (lattice results are unaltered). To compare
our results with those of the hotQCD Collaboration a new definition for the 
Polyakov loop was applied, thus a direct comparison with Refs. \cite{6} 
and \cite{7} is not possible. As we emphasized, the various $T_c$ 
values do not indicate separate phase transitions but the broadness of the
cross-over. Thus, it is more informative to look at the complete T dependence
of the observables (see the figures of this section) than just at the
definition-dependent characteristic points of them.  The
Bielefeld-Brookhaven-Columbia-Riken Collaboration \cite{1}  (independently of
the observables) obtained $T_c$=192(4)(7) for physical quark masses in the
continuum limit. The published results of the hotQCD Collaboration indicate a
narrow transition within the 185--195~MeV temperature range (for which they
expected about 5~MeV shift to smaller $T$ values in the continuum limit and
another 5~MeV because they used non-physical quark masses). Recent, preliminary
results of the hotQCD Collaboration move closer and closer to our curves,
and the original $\approx$40~MeV discrepancy in chiral variables 
is reduced to about 10~MeV (though the continuum extrapolated hotQCD result is
missing). For a detailed comparison of our and their results see the next
section.
}
\label{table2}
%\end{table}
}

In this section we presented our primary results, the temperature dependence of
various observables the same way as we did in our previous works \cite{6,7}. We
found a complete agreement.  For the readers' convenience we tabulate the
results in the Table~of Appendix \ref{B}. These curves contain the complete
information on the observables. Nevertheless, it is usual to determine some
characteristic points of these curves (inflection points or peaks). Since the
transition is a broad cross-over, these $T_c$ values scatter within the
transition range (c.f. Table~\ref{table2}, where we also review the results of
our previous analyses for comparison).

For completeness we discuss the trace anomaly too (see next section) and give
the transition temperature obtained from it and from the energy density in
Table \ref{table2} (the details of the equation of state at $N_t=6,8,10$ and 12
will be given in a subsequent publication \cite{EOS}).  The uncertainties are
given in parentheses. The first one refers to $T>0$, the second one to $T=0$
statistical plus systematic errors.  \section{Hadron Resonance Gas model
\label{hrg}} The Hadron Resonance Gas model has been widely used to study the
low temperature phase of QCD in comparison with lattice data
\cite{Karsch:2003vd,Karsch:2003zq,Tawfik:2004sw}.  In Ref.
\cite{Huovinen:2009yb} an important ingredient was included in this model,
namely the pion mass- and lattice spacing-dependence of the hadron masses. In
the present paper we combine these ingredients with Chiral Perturbation Theory
($\chi$PT)  \cite{14}. This opens the possibility to study chiral quantities,
too.

\subsection{The partition function of the HRG model}
The low temperature phase of QCD is dominated by pions. Goldstone's theorem 
implies weak interactions between pions at low energies, which allows to study them 
within $\chi$PT. As the temperature $T$ increases, 
heavier states become more relevant and need to be taken into account. The Hadron 
Resonance Gas model has its roots in the theorem by Dashen, Ma and Bernstein \cite{15}, 
which allows to calculate the microcanonical partition function of an interacting
system, in the thermodynamic limit $V\rightarrow\infty$, 
%as the sum of the free one, plus 
%a term depending only on the physical scattering matrix. It can be shown that, 
%if only the resonant part of the scattering matrix is retained and the background 
%interaction can be neglected, the second term is equivalent to considering all hadronic 
%resonances as free particles with distributed mass. For large enough temperatures, 
%the effective mass approaches the physical one: this picture holds if the energy 
%density or temperature of the system is large enough for most resonances to be 
%excited. Therefore, in this temperature regime an interacting hadron gas is, to a 
to a good approximation, assuming that it is a gas of non-interacting free hadrons and resonances \cite{16}. 
The pressure of the Hadron Resonance Gas model can be written as 
the sum of independent contributions coming from non-interacting resonances
\begin{eqnarray}
\frac{p^{HRG}}{T^4}
=\frac{1}{VT^3}\sum_{i\in\;mesons}\hspace{-3mm} 
\ln{Z}^{M}(T,V,\mu_{X^a},m_i)
+\frac{1}{VT^3}
\sum_{i\in\;baryons}\hspace{-3mm} \ln{Z}^{B}(T,V,\mu_{X^a},m_i)\; ,
\label{eq:ZHRG}
\end{eqnarray}
where
\begin{eqnarray}
\ln{Z}^{M}(T,V,\mu_{X^a},m_i)
&=&- \frac{V {d_i}}{{2\pi^2}} \int_0^\infty dk k^2
\ln(1- z_ie^{- \varepsilon_i/T}),
\nonumber\\
 \ln{ Z}^{B}(T,V,\mu_{X^a},m_i)
 &=&\frac{V {d_i}}{{2\pi^2}} \int_0^\infty dk\, k^2
\ln(1+ z_ie^{- \varepsilon_i/T})  ,
\label{eq:ZMB}
%\nonumber
\end{eqnarray}
with energies
$\varepsilon_i=\sqrt{k^2+m_i^2}$, degeneracy 
factors $d_i$ and fugacities
\begin{eqnarray}
z_i=\exp\left((\sum_a X_i^a\mu_{X^a})/T\right) \; .
\label{eq:fuga}
%\nonumber
\end{eqnarray}
In the above equation, $X^a$ are all possible conserved charges, including the baryon number $B$,
electric charge $Q$ and strangeness $S$. The sums in Eq. (\ref{eq:ZHRG}) include all
known baryons and mesons up to 2.5 GeV, as listed in the latest edition of the Particle 
Data Book \cite{17}. 

We will compare the results obtained with the physical hadron masses to those 
obtained with the distorted hadron spectrum which takes into account lattice discretization effects. As shown in Section \ref{taste}, each pseudoscalar meson in the staggered formulation is split into 16 mesons with different masses. The 
contribution of each meson to the pressure is given by: 
\bea
\frac{p^{\pi,K}}{T^4} 
=\frac{1}{16}\frac{1}{VT^3}\sum_{i=0}^{7}n_i 
\ln{Z}^{M}(T,V,\mu_{X^a},m_i)
\eea
where $m_i$ are the taste-split pseudoscalar meson masses (for the pion they are shown in
Fig. \ref{fig1}) and $n_i$ are the degeneracies listed in Table \ref{table1}.
Similarly to Ref. \cite{Huovinen:2009yb}, we will also take into account the pion mass- and lattice spacing- 
dependence of all other hadrons and resonances. 

%\begin{figure}
\FIGURE{
\centerline{\includegraphics[width=3in]{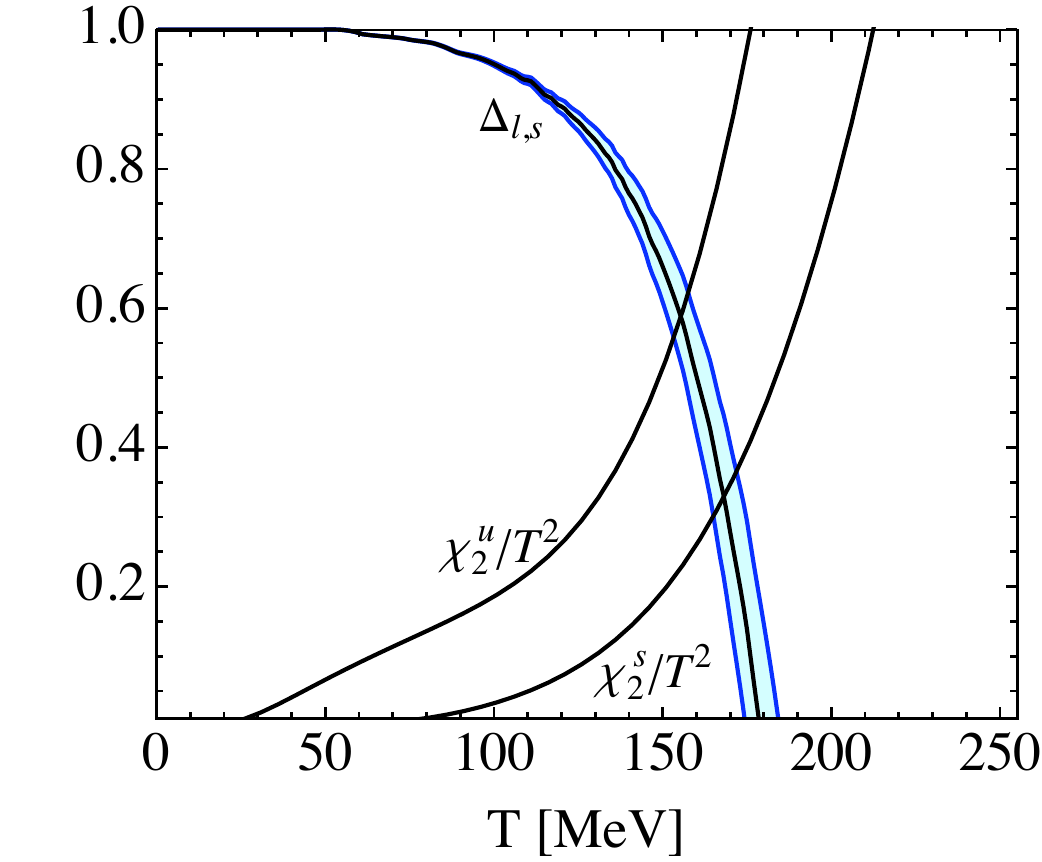}}
\caption{
HRG + $\chi$PT results for the light and strange quark number susceptibilities and 
the subtracted chiral condensate $\Delta_{l,s}$. For this last quantity, the error band 
indicates the uncertainty on the quark mass-dependence of hadrons, see the Appendix. 
The results have been obtained with physical values for the hadron and resonance 
masses, thus no lattice artefact  has been included.}
\label{fig2}
}
%\end{figure}

In order to calculate the chiral condensate in the HRG model, we need to know the behavior of all 
baryon and meson masses as functions of $m_l$ and $m_s$ . For the quark mass-dependence of 
the ground state hadrons, we use the most recent fits from $\chi$PT available in the literature
\cite{MartinCamalich:2010fp}. 
The same study is not available for all the resonances that we include. Therefore, 
similarly to Ref. \cite{Huovinen:2009yb}, we work under the assumption that all resonance masses behave 
as their fundamental states as functions of $m_q$. In addition, in order to have a more 
precise estimate, we determine the contribution of pions to the chiral condensate obtained 
in three-loop chiral perturbation theory in \cite{18}. All other hadrons and resonances 
are still treated in the ideal gas approximation. 
All details of this calculation are given in Appendix A. 
The HRG model + $\chi$PT results for light and strange quark number susceptibilities, and for the chiral condensate, are 
shown in Fig. \ref{fig2}.

It is instructive to look at these curves, before comparing them to 
the lattice results. In the low temperature phase, $\chi_{2}^{u}$ is dominated by pions, while 
$\chi_{2}^{s}$ by kaons; this is the reason why the light quark susceptibility rises much faster 
with increasing temperature, compared to the strange one. In the HRG model the 
susceptibilities keep increasing and $\Delta_{l,s}$ keeps decreasing to a negative value with 
increasing temperature. In QCD, all three quantities take values between 0 and 1. 
One can therefore take 0.5 as an illustration for the definition of $T_c$. From Fig.~\ref{fig2}, it is evident that one obtains 
similar transition temperatures for $\Delta_{l,s}$ and $\chi_{2}^{u}$, around 150 MeV, while 
$\chi_{2}^{s}$ reaches the 0.5 value at a larger temperature, around 170 MeV.
From this figure it is also evident that it is not the mere value of $T_c$ which is relevant in order
to describe the phase transition, but rather the full temperature dependence of the curves, from which it is immediately clear that different observables may produce very different values for the
transition temperature.

\subsection{Comparison between HRG model and lattice results \label{results}}
In this paper we compare two sets of lattice data:
\begin{itemize}
 \item{The first set is based on the Wuppertal-Budapest results.}
 \item{The second set is obtained by the Bielefeld-Brookhaven-Columbia-Riken
Collaboration, which later merged with a part of the the MILC collaboration 
and formed the hotQCD collaboration.} 
\end{itemize}
Furthermore, we use two types of 
theoretical description (based on hadron resonance gas model and 
chiral perturbation theory, for short: HRG+$\chi$PT):
\begin{itemize}
 \item{ One of the theoretical descriptions is based on the physical spectrum from
 the Particle Data Book
 (we call this description ``physical").}
 \item{ The other theoretical approach is based on a non-physical spectrum (this spectrum
 is obtained by $T=0$ simulations of the action one studies; the
 reason for this distortion will be explained later); we call this description ``distorted".}
\end{itemize}
As it is known, the Wuppertal-Budapest and the hotQCD results disagree. 
All characteristic temperatures are higher for the hotQCD Collaboration.
Note, that this discrepancy is not related to the difficulty of determining
e.g. inflection points of slowly varying functions (typical for a broad
cross-over). The discrepancy appears for all variables for a large
temperature interval. As we claimed earlier \cite{7} we observed 
``approximately 
20--35 MeV difference in the transition regime between our results and 
those of the hotQCD Collaboration".

As we will see, the Wuppertal-Budapest results are in complete agreement 
with the ``physical" hadron resonance gas model and with the ``physical" 
chiral perturbation theory, whereas the hotQCD results cannot be 
described this way. The hotQCD results can only be described by the 
``distorted" HRG+$\chi$PT.

%\begin{figure}
\FIGURE{
\hspace{-.8cm}
\begin{minipage}{.48\textwidth}
\parbox{6cm}{
\scalebox{.65}{
\includegraphics{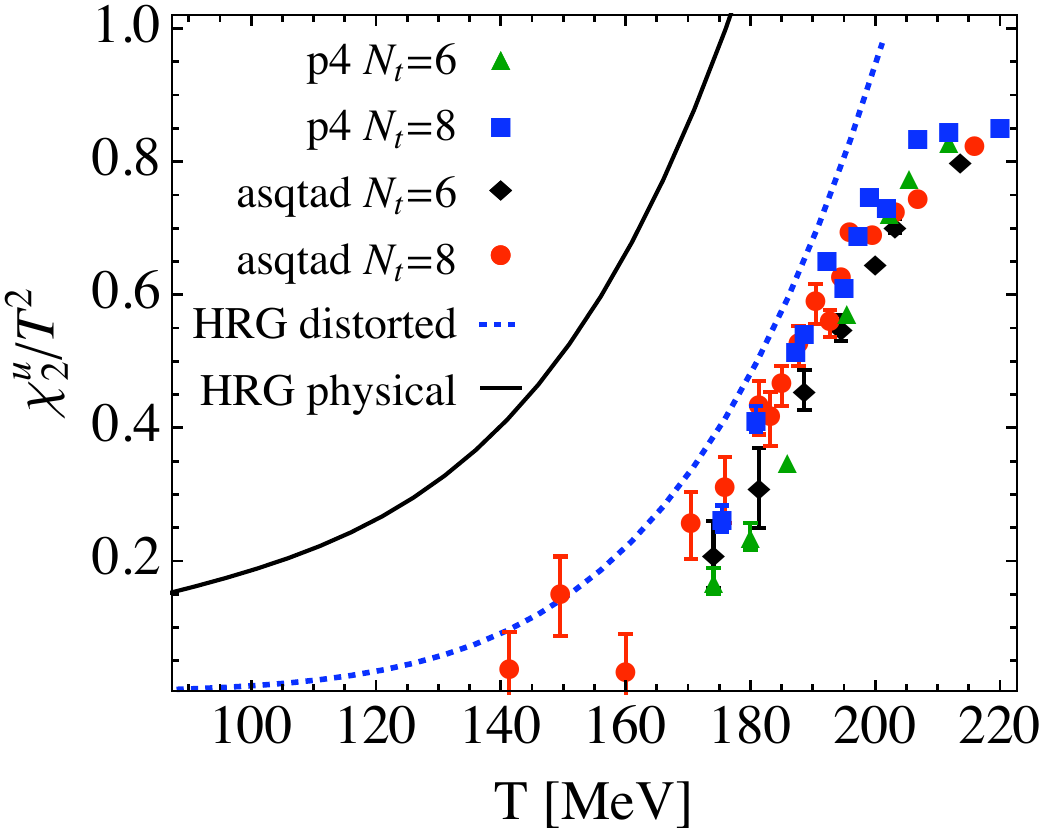}\\}}
%\centerline{(a)}
\end{minipage}
\hspace{.24cm}
\begin{minipage}{.48\textwidth}
\parbox{6cm}{
\scalebox{.65}{
\includegraphics{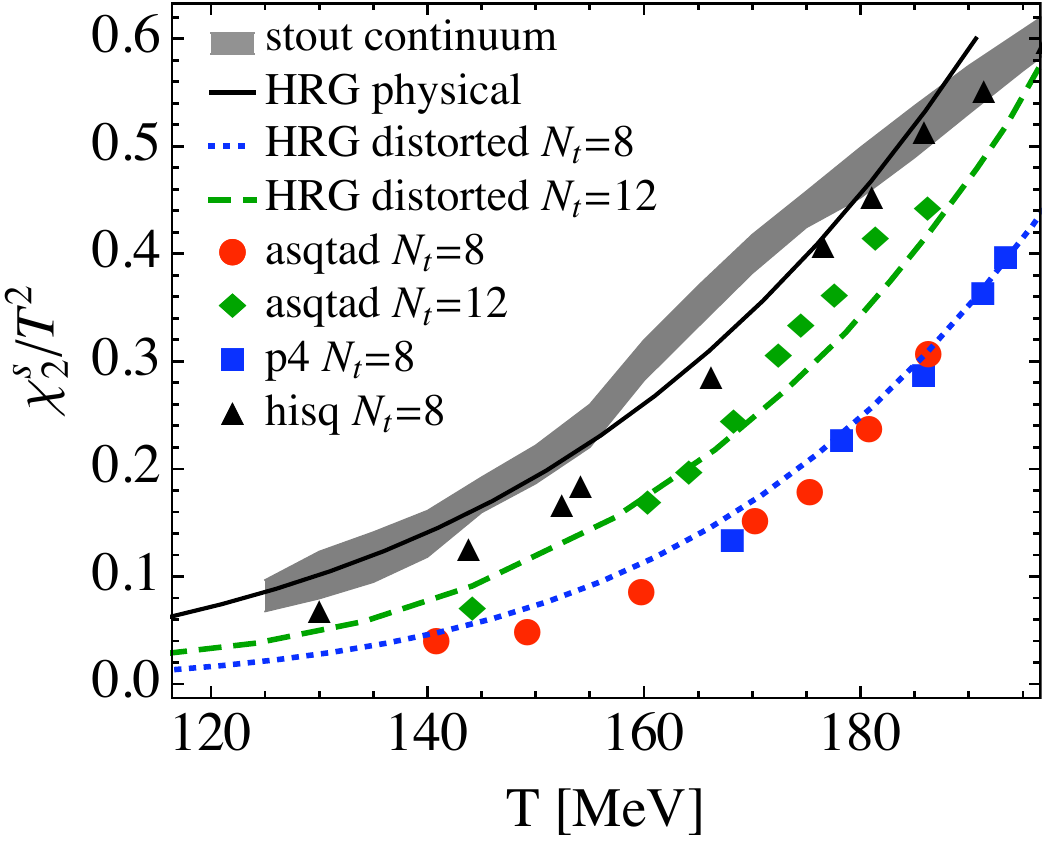}\\}}
%\centerline{(b)}
\end{minipage}
\caption{
Left: light quark susceptibility as a function of the temperature. Right: 
strange quark susceptibility as a function of the temperature. In both panels, the points 
with different symbols correspond to results obtained with the asqtad and p4
actions \cite{Bazavov:2009zn,Bazavov:2010sb}.  The solid line is the HRG model result with physical
masses. The dashed and dotted lines are the
HRG model results with distorted masses corresponding to $N_t=12$ and $N_t=8$, which take into account the discretization
effects and heavier quark masses, which characterizes the results of the hotQCD Collaboration.
Our continuum result for the strange susceptibility is shown by the band. 
Good agreement is found with the physical HRG+$\chi$PT results. (Due to 
the noisy contribution of 
the disconnected diagrams we
have not determined the light quark susceptibility.)}
\label{fig3}
}
%\end{figure}

In Fig.~\ref{fig3} we show the light and strange quark number susceptibilities, in the left and right panels, respectively. The lattice results are
compared to the HRG model predictions for physical (solid line) and distorted (dashed line) spectrum 
(due to the noisy contribution of the disconnected diagrams we don't have results for the light quark susceptibility).
The distorted spectrum takes 
into account the larger quark masses used by the hotQCD collaboration, as well 
as the larger lattice spacing and pseudoscalar meson splittings (see figure
\ref{fig1}). For all hadrons and resonances, we use the pion mass- and lattice
spacing-dependence given in Refs. \cite{Bazavov:2009bb,masses} and parametrized in Ref.
\cite{Huovinen:2009yb}. As we can see, once we take these effects into account
(which corresponds to a distorted spectrum), the HRG curves on both figures are
sensibly different from the physical ones and agree with the corresponding
lattice data of hotQCD.  Our continuum results on the strange susceptibility are
compared to the other results, too.  We observe a good agreement between our
results and the ``physical'' HRG ones\footnote{For completeness we included in
our comparisons preliminary \cite{Bazavov:2010sb} results of the hotQCD
collaboration obtained by the HISQ and asqtad actions on $N_t$=8 and 12,
respectively. We will discuss their impact
later.}. Notice that  the agreement between lattice and HRG model results is good below the transition temperature, while for larger temperatures a deviation is obviously expected. This is observed both in our results and the hotQCD ones, but the temperatures at which deviations occur are obviously different.
%Including the distortion obtained from our stout action at our lattice spacings results in HRG %curves, which are in the close vicinity of the physical HRG curve.
\FIGURE{
%\begin{figure}
\hspace{-.8cm}
\begin{minipage}{.48\textwidth}
\parbox{6cm}{
\scalebox{.65}{
\includegraphics{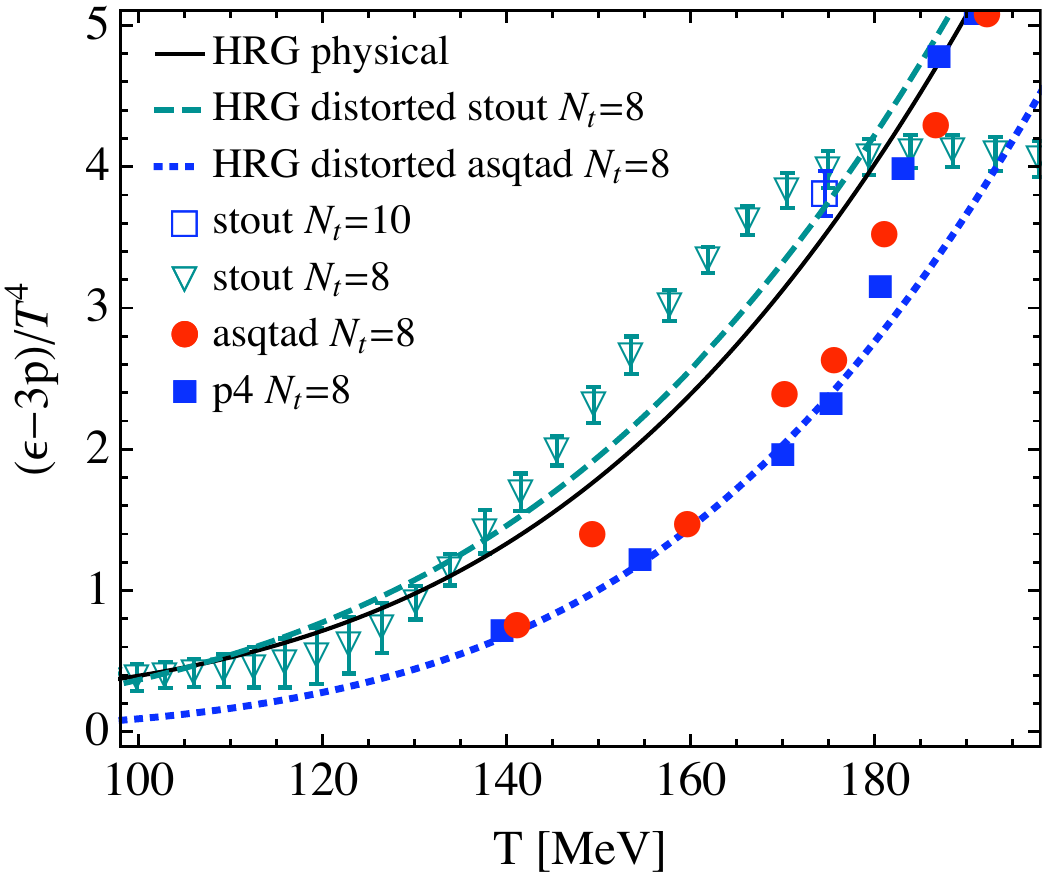}\\}}
%\centerline{(a)}
\end{minipage}
\hspace{.24cm}
\begin{minipage}{.48\textwidth}
\parbox{6cm}{
\scalebox{.65}{
\includegraphics{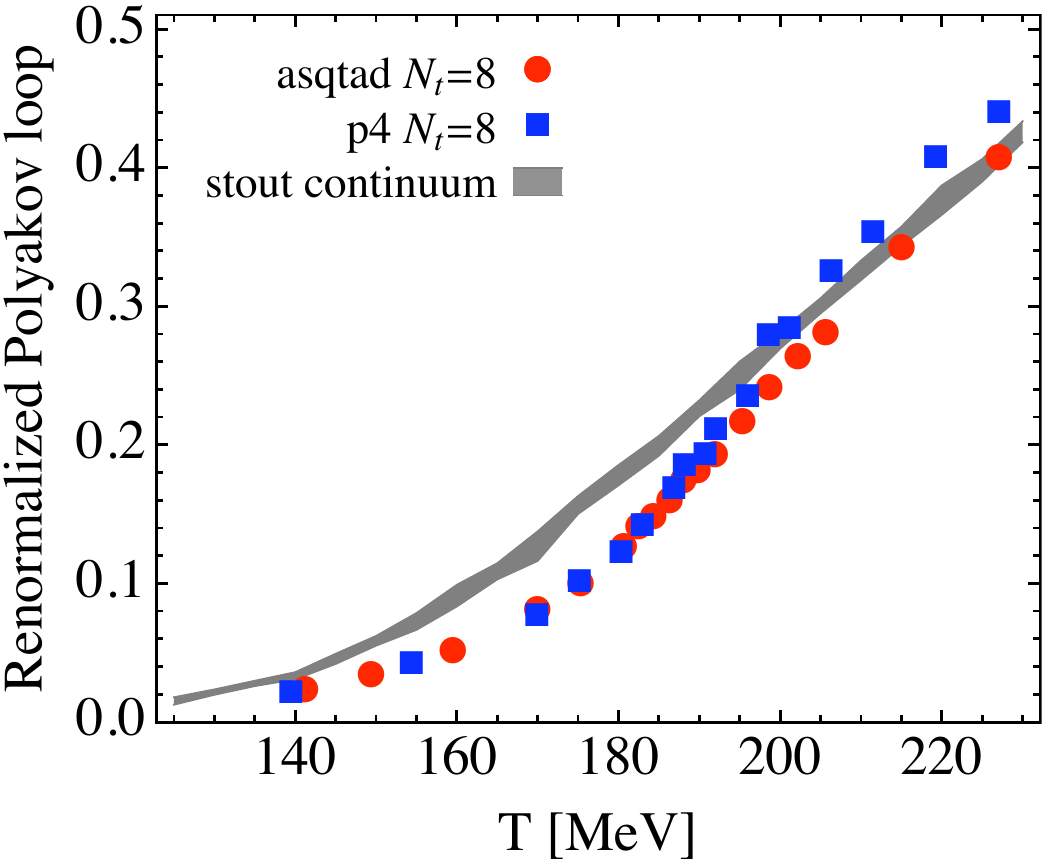}\\}}
%\centerline{(b)}
\end{minipage}
\caption{
Left: $(\epsilon-3p)/T^4$ as a function of the temperature. Open symbols are our results. Full symbols are the results for the asqtad and p4 actions at $N_t=8$ \cite{Bazavov:2009zn}. Solid line: HRG model with physical masses. Dashed lines: HRG model with distorted spectrums. As it can be seen, the prediction of the HRG model with a spectrum distortion corresponding to the stout action
at $N_t=8$ is already quite close to the physical one. The error on the recent preliminary HISQ result \cite{Bazavov:2010sb} is larger than the difference between the stout and asqtad data, that is why we do not show them here. Right: renormalized Polyakov loop. We compare our results with those of the hotQCD Collaboration  (asqtad and p4 data for $N_t=8$ \cite{Bazavov:2009zn}).
 }
\label{fig6}
%\end{figure}
}

In the left panel of Fig. \ref{fig6} we show the trace anomaly divided by $T^4$ as a function of
the temperature. Our $N_t=8$ results are taken from Ref. \cite{EOS}. Notice
that, for this observable, we have a check-point at $N_t=10$ too: the results
are on top of each other. Also shown are the results
of the hotQCD collaboration at $N_t=8$ \cite{Bazavov:2009zn} and the HRG model predictions for physical and distorted resonance spectrums. On the one hand, our results are in good agreement with the ``physical'' HRG model ones. It is important to note, that using our mass splittings and inserting this  distorted spectrum into the HRG model gives a temperature dependence which lies essentially on the physical HRG curve (at least within our accuracy). On the other hand, a distorted spectrum based on the asqtad and p4 frameworks results in a shift of about 20 MeV to the right. The asqtad and p4 lattice results can be successfully described by this distorted HRG prediction, too. 

In the right panel of Fig. \ref{fig6} we show the renormalized Polyakov loop
(the renormalization procedure was discussed in the previous Section). The
comparison with the data of \cite{Bazavov:2009zn} shows a good agreement at
high temperatures, and deviations in the transition region.

In Fig. \ref{fig4}, we show results for the chiral condensate as a function of
the temperature.  The left panel shows $\langle\bar{\psi}\psi\rangle_R$ as
defined in Eq. (\ref{pbp}), while the right panel shows $\Delta_{l,s}$ (see Eq.
(\ref{deltals})).  In both panels, the solid black curve has been obtained in
the HRG+$\chi$PT model, using the equations given in Appendix A. The error
bands of the theoretical lines correspond to the uncertainty in the quark mass
dependence of hadron masses \cite{MartinCamalich:2010fp}. Gray bands correspond
to our continuum results. They agree with the ``physical'' HRG+$\chi$PT
predictions. In the right panel, we also show the lattice results for the
subtracted chiral condensate obtained with the asqtad and p4 actions
\cite{Bazavov:2009zn,Bazavov:2010sb}. These results are compared to the dashed
curves, which have been obtained in the HRG+$\chi$PT model with ``distorted''
masses corresponding to $N_t=8$ and $N_t=12$.  Also in this case, for all
hadrons and resonances we use the pion mass- and lattice 
spacing-dependence taken from Refs. \cite{Bazavov:2009bb,masses} and parametrized in Ref. \cite{Huovinen:2009yb}. 
%To complete our analysis, we have 
%calculated the subtracted chiral condensate in the HRG+$\chi$PT model with a spectrum that 
%takes into account the lattice discretization effects and pseudoscalar meson splitting
%corresponding to our stout action with $N_t = 16$. The curve lies within the theoretical error 
%bands shown in Fig. \ref{fig4}. 

From all quantities that we have calculated, a consistent picture arises: our
stout results agree with the ``physical'' HRG+$\chi$PT predictions; whereas the
observed shift in transition temperatures between the results of the stout and
the asqtad and p4 actions can be easily explained within the Hadron Resonance
Gas+$\chi$PT model with ``distorted'' masses. Once the discretization effects,
the taste violation and the heavier quark masses used in
\cite{Bazavov:2009zn,Bazavov:2010sb} are taken into account, all the
HRG+$\chi$PT curves for the different physical observables are shifted to
higher temperatures and fall on the corresponding lattice results. 

As a final check, we have determined the chiral condensate with larger quark masses ($m_s/m_{u,d}=3$, corresponding to 
a pseudo-Goldstone mass of about 414~MeV and to an average pion mass of 587 MeV, which matches the one of Ref. \cite{Bazavov:2009zn}
at $a=0.183$ fm, corresponding to the lower end of the transition region
$T=135$ MeV at $N_t=8$). Notice that, due to the reduced taste splitting of the stout action, we need $m_s/m_{u,d}=3$ in order to have an average pion mass compatible with the one of Ref. \cite{Bazavov:2009zn}, where $m_s/m_{u,d}=10$.
The results of this run are shown in Fig. \ref{chiheavy}. This procedure allows us to reproduce the results of the hotQCD collaboration for this observable, though with an artificially large quark mass. This example illustrates that a large pion splitting (of the asqtad action) results in a physically distorted spectrum, which can be mimicked by a small splitting (of the stout action) at an artificially large quark mass.
\FIGURE{
%\begin{figure}
\hspace{-.8cm}
\begin{minipage}{.48\textwidth}
\parbox{6cm}{
\scalebox{.65}{
\includegraphics{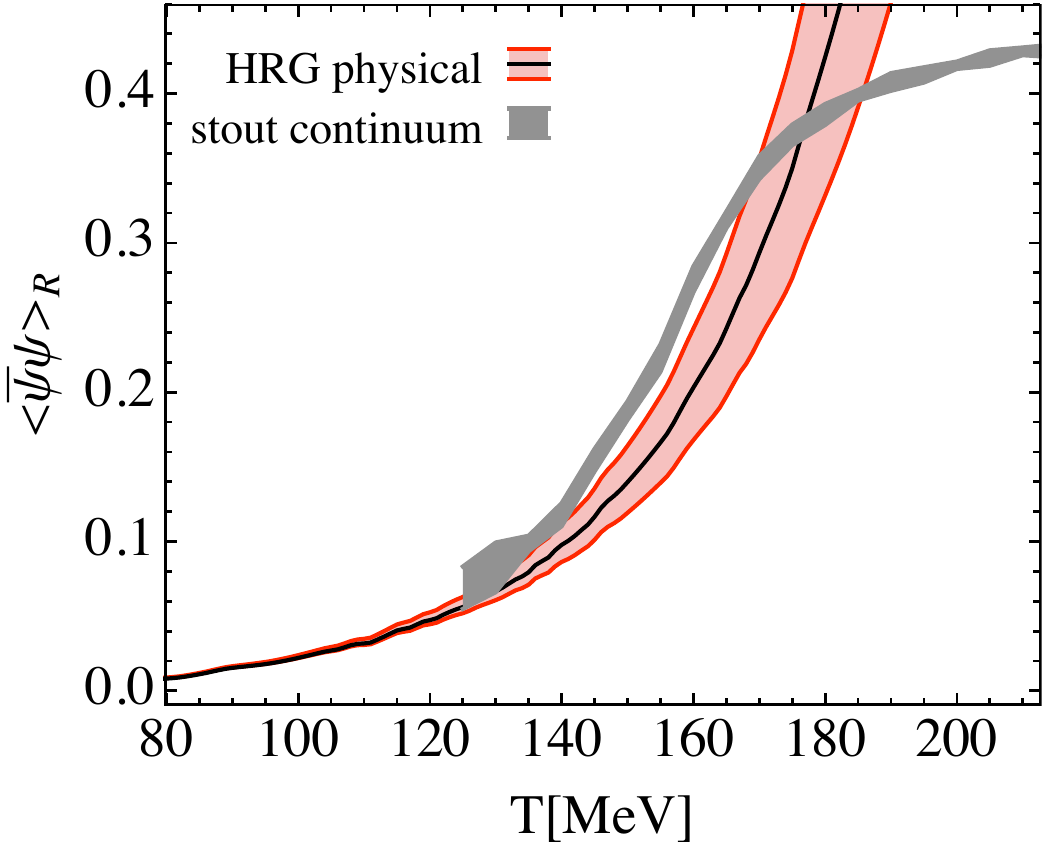}\\}}
%\centerline{(a)}
\end{minipage}
\hspace{.24cm}
\begin{minipage}{.48\textwidth}
\parbox{6cm}{
\scalebox{.65}{
\includegraphics{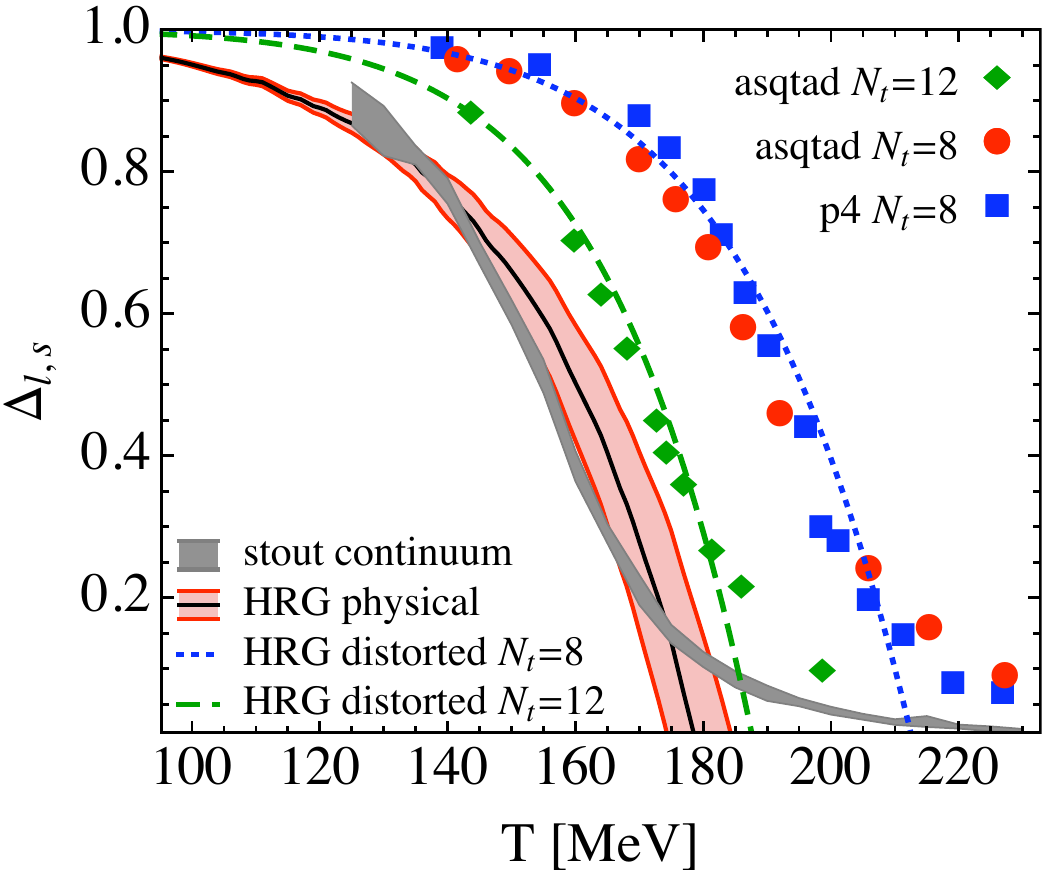}\\}}
%\centerline{(b)}
\end{minipage}
\caption{
Left: Renormalized chiral condensate as defined in Eq. (\ref{pbp}). Right: Subtracted chiral condensate $\Delta_{l,s}$ as defined in eq. (\ref{deltals}), as a function of the 
temperature. Gray bands are the continuum results of our collaboration, obtained with the 
stout action. Full symbols are obtained with the asqtad 
and p4 actions \cite{Bazavov:2009zn,Bazavov:2010sb}. In both panels, the solid line is the HRG model result with physical masses. 
The error band corresponds to the uncertainty in the quark mass-dependence of 
hadron masses. The dashed lines are the HRG+$\chi$PT model result with distorted masses, 
which take into account the discretization effects and heavier quark masses used in 
\cite{Bazavov:2009zn,Bazavov:2010sb} for $N_t=8$ and $N_t=12$.
}
\label{fig4}
%\end{figure}
}

\FIGURE{
%\begin{figure}
\centerline{\includegraphics[width=3in]{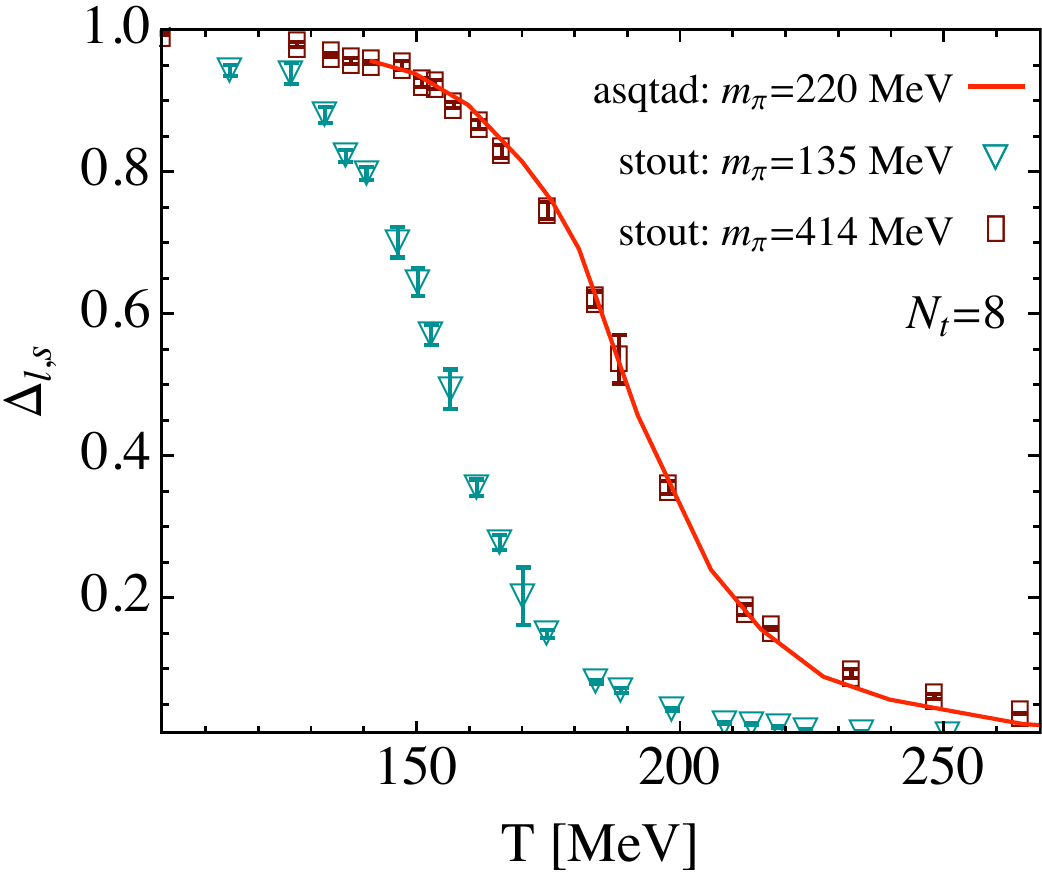}}
\caption{
Subtracted chiral condensate $\Delta_{l,s}$ as a function of the temperature. The empty triangles are our results with physical quark masses as shown in Fig. \ref{chir}. The empty rectangles are our results with an average pion mass of 587 MeV at $T\simeq135$ MeV. The red curve is the result of the hotQCD collaboration \cite{Bazavov:2009zn}: these results are the same shown in Fig. \ref{fig4}: a line connects the data to lead the eye. For all sets of data we have $N_t=8$. As it can be seen, the asqtad data can be mimicked in the stout framework by using a larger quark mass.}
\label{chiheavy}
}
%\end{figure}

There is a proceedings contribution written by two members of the hotQCD 
Collaboration, in which the HISQ action is applied \cite{20} and 
preliminary results are presented. This action uses a highly improved smearing recipe (and similarly to our
stout action it reduces the pion splitting much more than the asqtad
or p4 actions). In contrast to previous findings of the hotQCD 
collaboration, the results based on this new smeared improved action are 
quite close to our results. Both the strange susceptibility and the 
chiral condensate curves shift to lower temperatures. The approximately 
20 MeV discrepancy for the strange susceptibility between the 
Wuppertal-Budapest and the hotQCD results has essentially disappeared. 
The approximately 35 MeV discrepancy for the chiral condensate curves is 
reduced to about 10 MeV (see Fig. \ref{fig5}). 
One expects that the results with the HISQ action 
will approach the continuum results much faster than those with the previously applied
asqtad or p4 actions of the hotQCD collaboration. Note, that the continuum limit within the HISQ 
framework is still missing. This last important step (which needs quite 
some computational resources and also care) will hopefully eliminate the 
remaining minor discrepancy, too. 
The same two members of the hotQCD
Collaboration presented preliminary results using the asqtad action on
$N_t$=12 lattices \cite{Bazavov:2010sb}, too. 
At this lattice spacing the pion splitting is smaller
than on $N_t$=8 lattices, and the curves move closer to ours. Since
this action and lattice spacing combination has still a 
larger splitting than the HISQ result, it is further away from
our continuum results than the findings within the HISQ framework.
Following these two authors (Figure 5. of Ref. \cite{Bazavov:2010sb}) we zoom in into the transition
region of $\Delta_{l,s}$ and on Figure~\ref{fig5} we show various 
findings. The stout results from a broad range of lattice spacings
($N_t$=8, 10, 12 and 16) are shown with open symbols. They all accumulated
in the vicinity of our continuum estimate, indicated by the thin gray band.
The hotQCD results were obtained by three different actions (p4, asqtad
and HISQ) and with two different pion masses (220 and 160~MeV). They
cover a broad range. The smaller the pion mass and/or pion splitting
in the hotQCD results, the closer it is to ours.

These results confirm the expectations 
\cite{6,7} that the source of the discrepancy was 
the lack of the proper continuum extrapolation \cite{6} in the 
hotQCD result: a dominant discretization artefact within the asqtad 
and p4 actions is the large pion splitting \cite{Fodor:2007sy}, 
which resulted in the distorted spectrum. 

 As we emphasized in both our 
previous studies \cite{6,7}, only continuum 
extrapolated results are physical. We demonstrated \cite{6} 
that using $f_K$ and $r_0$ scale settings gives the same continuum 
result. Furthermore, we showed that using other quantities (the masses 
of the $\Omega$, K$^*$ and $\Phi$ hadrons or the pion decay constant) the 
same continuum result is obtained \cite{7}. In this sense 
(thus after continuum extrapolation) one scale setting can be 
susbstituted by another one, the result remains the same. 
%The proper, but quite expensive procedure, which will probably be carried out,
%is to perform continuum extrapolation with the HISQ action, too. 
The ideal situation would be to compare our (continuum) results and the results obtained
by continuum extrapolation based on HISQ (asqtad, p4 or any other) action.
Without continuum extrapolation, cutoff effects appear, which can manifest themselves
by providing different scales from different observables.
As it is now,
for the chiral condensate there is about 10 MeV difference between the
continuum result of the Wuppertal-Budapest collaboration and the non-continuum
$N_t$=8 results obtained by the HISQ action (Note, that this is much smaller
than previous findings of the hotQCD Collaboration indicated). As we mentioned,
carrying out the continuum extrapolation with the HISQ action 
will probably remove even this small difference.

\FIGURE{
%\begin{figure}
\centerline{\includegraphics[width=4in]{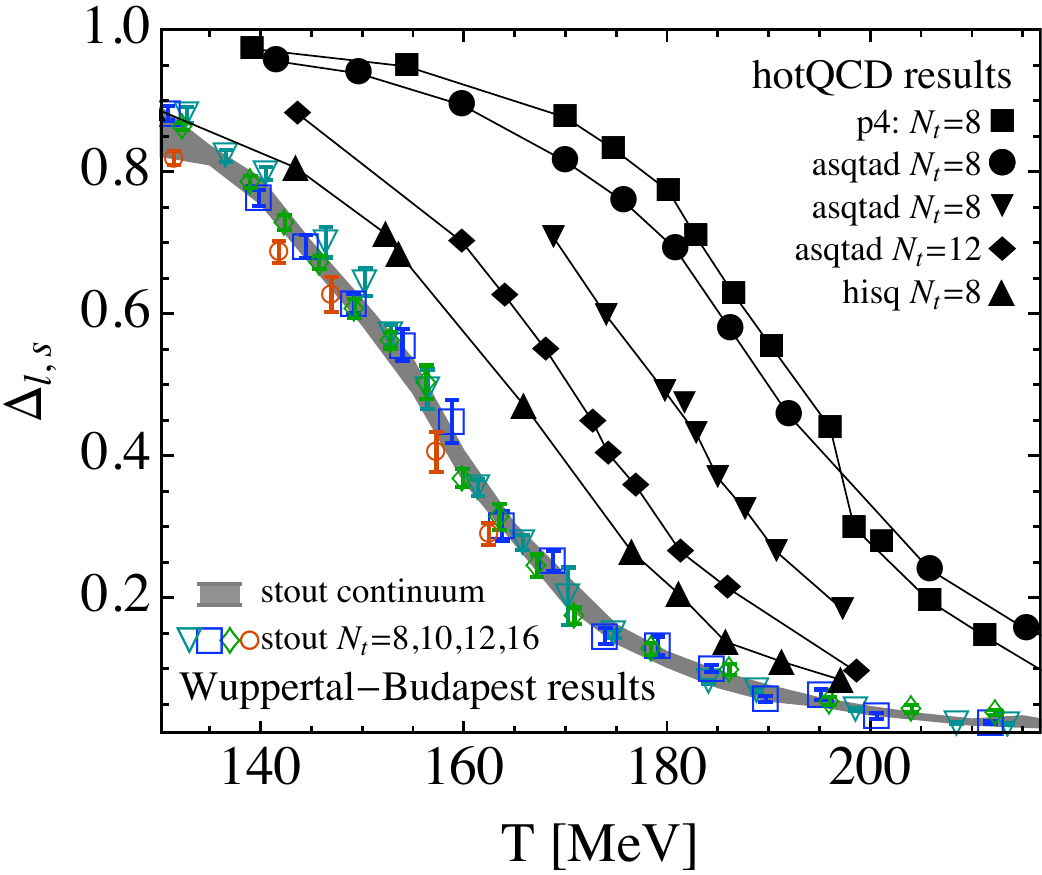}}
\caption{
The subtracted chiral condensate $\Delta_{l,s}$ as a function of the
temperature.  We show a comparison between stout, asqtad, p4 and
HISQ \cite{Bazavov:2009zn,Bazavov:2010sb} results. Our results are shown by
colored open symbols, whereas the hotQCD results are shown by full black
symbols. The gray band is our continuum result, the thin lines for the hotQCD
data are intended to lead the eye. Our stout results were all obtained by the
physical pion mass of 135 MeV. The full dots and squares correspond to
$m_\pi=220$ MeV, the full triangles and diamonds correspond to $m_\pi=160$ MeV
of the hotQCD collaboration.  }
\label{fig5}
%\end{figure}
}

\section{Conclusions\label{conclusions}}
We have presented our latest results for the QCD transition temperature. The 
quantities that we have studied are the strange quark number susceptibility, the Polyakov loop,
the chiral condensate and the trace anomaly. We have given the complete
temperature dependence of these quantities, which provide more
information that the characteristic temperature values alone.
Our previous results for the strange quark susceptibility, the Polyakov loop and 
the chiral condensate have been pushed to an even finer lattice ($N_t = 16$).
The new data corresponding 
to $N_t = 16$ confirm our previous results. The trace anomaly \cite{EOS} was obtained for $N_t=8$ and a check-point at $N_t=10$. The transition temperature that we obtain from this last quantity is very 
close to the one obtained from the chiral condensate. 

In order to find the origin of the discrepancy between the results of our collaboration and the hotQCD ones, we calculated these observables (except the Polyakov loop) in the Hadron 
Resonance Gas model. Besides using the physical hadron masses, we also
performed the calculation with modified masses which take into account the
heavier pions and larger lattice spacings used in \cite{Bazavov:2009zn}. We
find an agreement between our data and the HRG ones with ``physical'' masses,
while the hotQCD collaboration results are in agreement with the HRG model only
if the spectrum is ``distorted'' as it was directly measured on the lattice
\cite{Bazavov:2009bb,masses}. This analysis therefore provides an easy and convincing
explanation of the observed shift in transition temperatures between the two
collaborations and emphasizes the role of the proper continuum limit.

We used 2+1 flavor QCD within the staggered framework, which needs taking the root of the fermion determinant.
There is a lively discussion in the literature whether this is a correct procedure. Though we have
not seen any problem with this fermion formalism (our results and the predictions of the
hadron resonance gas model agree very nicely up to the transition region) it is still very important
to repeat the calculations with actions, which are free of the rooting problem (e.g. Wilson fermions).
\section*{Acknowledgements}
Computations were performed on the Blue Gene supercomputers at FZ 
J\"ulich and on clusters at Wuppertal and also at the E\"otv\"os  
University, Budapest. This work is supported in part by European Union  
(EU) grant I3HP; Deutsche Forshungsgemeinschaft grants FO 502/2 and  
SFB-TR 55 and  
(FP7/2007-2013)/ERC no. 208740, and by
the U.S. Department of Energy under Grant No. DE-FG02-05ER25681. The authors also thank Peter Petreckzy for  
the information on the preliminary results by the HotQCD
Collaboration, and Stefan D\"urr for stimulating discussions.
\section*{Appendix}
\appendix
\section{Renormalized chiral condensate}
\renewcommand{\theequation}{A-\arabic{equation}}
  \setcounter{equation}{0}  
In order to calculate the subtracted chiral condensate $\Delta_{l,s}$ as defined in Eq. (\ref{deltals}), we need to calculate $\langle\bar{\psi}\psi\rangle_{u,T}$ and $\langle\bar{\psi}\psi\rangle_{s,T}$. The light quark chiral condensate is given 
by: 
\bea
\langle\bar{\psi}\psi\rangle_{u,T}&=&\langle\bar{\psi}\psi\rangle_{u,0}+\langle\bar{\psi}\psi\rangle_{\pi}+\frac{T}{V}\left[\sum_{i\in mesons}\frac{\partial\ln Z^M\left(T,V,\mu_{X^a},m_i\right)}{\partial m_i}\frac{\partial m_i}{\partial m_{\pi}^{2}}\frac{\partial m_{\pi}^{2}}{\partial m_u}\right.
\nonumber\\
&+&\left.\sum_{i\in baryons}\frac{\partial\ln Z^B\left(T,V,\mu_{X^a},m_i\right)}{\partial m_i}\frac{\partial m_i}{\partial m_{\pi}^{2}}\frac{\partial m_{\pi}^{2}}{\partial m_u}\right].
\label{chiu}
\eea
In the above formula, $\langle\bar{\psi}\psi\rangle_{u,0}$ is the chiral condensate at vanishing temperature, 
$\langle\bar{\psi}\psi\rangle_{\pi}$ is the temperature-dependent 
 pion contribution, that we take from the $\chi$PT investigation of Ref. \cite{18}. The sum over mesons in the 
square brackets obviously does not include pions. The derivatives of the hadron 
masses with respect to $m_{\pi}^{2}$ can be written as: 
\beq
\frac{\partial m_i}{\partial m_{\pi}^{2}}=\frac{\sigma_i}{(m_{\pi}^{2})_{phys}}
\eeq
where the $\sigma_i$ are the sigma terms evaluated at the physical pion mass. We use for our analysis the results recently obtained in \cite{MartinCamalich:2010fp} for the fundamental states. They agree with the results obtained by our collaboration in \cite{19}. We assume here that the resonances 
have the same sigma terms as their fundamental states. 
%\begin{table}
%\begin{center}
%\begin{tabular}{c||c|c}
%hadron&$\sigma_i$ [MeV]&error [MeV]\\
%\hline
%$\Xi$&12.46&7.19\\
%$\rho$& 31.62&28.62\\
%$K^*$&-0.50&12.78\\
%$\phi$&-3.42&4.39\\
%$N$&47.31&16.82\\
%$\Sigma$&31.68&10.53\\
%$\Delta$&88.00&52.09\\
%$\Omega$&12.13&10.90\\
%$\Lambda$&23.94&10.70\\
%\hline
%\end{tabular}
%\end{center}
%\caption{Sigma terms of the ground state hadrons.}
%\label{table3}
%\end{table}
Using the notations of Ref. \cite{18} we have: 
\beq
\langle\bar{\psi}\psi\rangle_{u,0}=\frac{m_{\pi}^{2}}{2m_u}\frac{F^2}{c}~~~~~~\Rightarrow~~~~~~
m_{\pi}^{2}=2\frac{\langle\bar{\psi}\psi\rangle_{u,0} c m_u}{F^2}~~~~~~\Rightarrow~~~~~~
\frac{\partial m_{\pi}^{2}}{\partial m_u}=2\frac{\langle\bar{\psi}\psi\rangle_{u,0}c}{F^2}
\eeq
In the above formulas, $c$ is a temperature-independent constant which is equal to 1 in the massless theory. The corrections of order $m_q$ have been calculated and give
\beq
c=0.90\pm 0.05;
\eeq
$F$ is the pion decay constant in the chiral limit:
\beq
F=88.3\pm 1.1 \mathrm{MeV}.
\eeq
Therefore, replacing the above relations in Eq. (\ref{chiu}), we obtain:
\bea
\langle\bar{\psi}\psi\rangle_{u,T}&=&\langle\bar{\psi}\psi\rangle_{u,0}\left\{1+\frac{\langle\bar{\psi}\psi\rangle_{\pi}}{\langle\bar{\psi}\psi\rangle_{u,0}}+2\frac{c}{F^2}\frac{T}{V}\left[\sum_{i\in mesons}\frac{\partial\ln Z^M\left(T,V,\mu_{X^a},m_i\right)}{\partial m_i}\frac{\sigma_i}{(m_{\pi}^{2})_{phys}}\right.\right.
\nonumber\\
&+&\left.\left.\sum_{i\in baryons}\frac{\partial\ln Z^B\left(T,V,\mu_{X^a},m_i\right)}{\partial m_i}
\frac{\sigma_i}{(m_{\pi}^{2})_{phys}}\right]\right\}
\label{chiu2}
\eea
We now need to calculate the strange quark condensate, $\langle\bar{\psi}\psi\rangle_{s,T}$. We proceed in a similar way as for the light quark condensate:
\bea
\langle\bar{\psi}\psi\rangle_{s,T}&=&\langle\bar{\psi}\psi\rangle_{s,0}+\langle\bar{\psi}\psi\rangle_{K}+\frac{T}{V}\left[\sum_{i\in mesons}\frac{\partial\ln Z^M\left(T,V,\mu_{X^a},m_i\right)}{\partial m_i}\frac{\partial m_i}{\partial m_s}\right.
\nonumber\\
&+&\left.\sum_{i\in baryons}\frac{\partial\ln Z^B\left(T,V,\mu_{X^a},m_i\right)}{\partial m_i}\frac{\partial m_i}{\partial m_s}\right].
\label{chis}
\eea
 $\langle\bar{\psi}\psi\rangle_{s,0}$ is the zero-temperature value of the strange condensate, which is related to $\langle\bar{\psi}\psi\rangle_{u,0}$ by QCD sum rules \cite{21}:
 \beq
 \langle\bar{\psi}\psi\rangle_{s,0}=0.8\langle\bar{\psi}\psi\rangle_{u,0}
 \eeq
$ \langle\bar{\psi}\psi\rangle_{K}$ is the kaon contribution to the strange condensate:
\beq
\langle\bar{\psi}\psi\rangle_{K}=\frac{\partial\ln Z^M\left(T,V,\mu_{X^a},m_K\right)}{\partial m_K}\frac{\langle\bar{\psi}\psi\rangle_{u,0}c}{2m_KF^2}.
\eeq
The strange mass dependence of hadrons and resonances can be written as:
\beq
\frac{\partial m_i}{\partial m_s}=\frac{\sigma_{i,s}}{m_s}=\sigma_{i,s}\frac{\langle\bar{\psi}\psi\rangle_{u,0}c}{m_{K}^{2}F^2}\frac{m_u+m_s}{m_s};
\eeq
the sigma terms $\sigma_{i,s}$ involving strange quarks are taken from Ref. \cite{MartinCamalich:2010fp}.
The sum over mesons in the square brackets of Eq. (\ref{chis}) does not include kaons. 

The ratio $(m_u + m_s )/m_s$ is equal to 29.15/28.15 for our collaboration, and to 11/10 or 21/20
for the hotQCD collaboration. 
We therefore have:
\bea
\langle\bar{\psi}\psi\rangle_{s,T}&=&\langle\bar{\psi}\psi\rangle_{u,0}\left\{0.8+\frac{\partial\ln Z^M\left(T,V,\mu_{X^a},m_i\right)}{\partial m_K}\frac{c}{2m_KF^2}\right.
\nonumber\\
&+&\left.\frac{T}{V}
\frac{c}{m_{K}^{2}F^2}\frac{m_u+m_s}{m_s}\left[\sum_{i\in mesons}\frac{\partial\ln Z^M\left(T,V,\mu_{X^a},m_i\right)}{\partial m_i}\sigma_{i,s}\right.\right.
\nonumber\\
&+&\left.\left.\sum_{i\in baryons}\frac{\partial\ln Z^B\left(T,V,\mu_{X^a},m_i\right)}{\partial m_i}
\sigma_{i,s}\right]\right\}
\label{chis2}
\eea

\section{Continuum results\label{B}}

In this paper we presented lattice data with $N_t=8,10,12$ and 16. Our continuum
extrapolation is based on these resolutions assuming a $\sim 1/N_t^2$ behaviour.
We used a fitted spline interpolation on the data. 
We summarize the results in Table~\ref{tab:curve}. 

In order to determine the transition temperatures we followed Ref.
\cite{19} and applied a combined 
fitting method weighting among various scenarios. The result is
a robust estimate. 
Note that the inflection points obtained using this method do not
necesserily agree with the inflection point of the mean values of 
Table~\ref{tab:curve}. Clearly, in an almost straight band (T 
dependent results with error bars) one can draw various curves with
different inflection points. That is the reason, why we emphasize 
more the complete temperature dependence than the individual
$T_c$ values. 

\TABLE{
%\begin{table}
%\begin{center}
\input{curvetab.tex}
%\end{center}
\caption{\label{tab:curve}
The continuum behaviour of our observables in the transition region
(please note that there is an uncertainty of about 2\% in the temperature
values corresponding to the systematics of setting the scale).
}
%\end{table}
}

\end{document}

%% file: curvetab.tex
\begin{tabular}{|c||c|c|c|c|}
\hline
T [MeV]&$\chi_2^s/T^2$&$L_{\rm Polyakov}$&$\langle\bar\psi\psi\rangle_R$&$\Delta_{l,s}$\\
\hline
125 & 0.08(1) & 0.015(3) & 0.07(1) & 0.89(3)\\
130 & 0.10(2) & 0.022(2) & 0.08(2) & 0.85(4)\\
135 & 0.12(2) & 0.028(2) & 0.099(5) & 0.81(1)\\
140 & 0.14(2) & 0.033(3) & 0.118(8) & 0.76(2)\\
145 & 0.18(2) & 0.045(4) & 0.155(8) & 0.67(2)\\
150 & 0.20(2) & 0.059(4) & 0.188(6) & 0.59(1)\\
155 & 0.24(2) & 0.073(6) & 0.223(9) & 0.49(3)\\
160 & 0.30(2) & 0.091(8) & 0.276(9) & 0.37(2)\\
165 & 0.35(2) & 0.109(6) & 0.315(6) & 0.28(1)\\
170 & 0.40(2) & 0.13(1) & 0.350(8) & 0.20(2)\\
175 & 0.44(2) & 0.157(7) & 0.372(7) & 0.14(1)\\
180 & 0.48(2) & 0.178(7) & 0.386(7) & 0.11(1)\\
185 & 0.51(2) & 0.199(7) & 0.399(4) & 0.08(1)\\
190 & 0.55(2) & 0.226(6) & 0.408(6) & 0.063(9)\\
195 & 0.59(2) & 0.25(1) & 0.413(5) & 0.051(5)\\
200 & 0.63(2) & 0.276(6) & 0.419(3) & 0.039(5)\\
205 & 0.65(2) & 0.300(6) & 0.424(5) & 0.031(4)\\
210 & 0.68(2) & 0.326(7) & 0.428(3) & 0.024(4)\\
215 & 0.70(2) & 0.350(7) & 0.429(4) & 0.023(4)\\
220 & 0.73(2) & 0.38(1) & 0.433(3) & 0.018(3)\\
\hline
\end{tabular}